\documentclass[ACS,STIX2COL]{WileyNJD-v2}
\articletype{Article Type}%

\received{}
\revised{}
\accepted{}

\usepackage{StyleBTO}
  \def\ele{160~\textcelsius }

\begin{document}

\title{Overcoming Discharge Inhibition in \textit{n}-Butane Oxidation: Two-Component \ch{BaTiO3} and Mn-Cu Mixed Oxide Coatings}

\author[1]{Timothy Oppotsch} 
\author[1]{Christian Oberste-Beulmann}
\author[2]{Alexander Böddecker} 
\author[2]{Gerrit Hübner} 
\author[2]{Ihor Korolov} 
\author[2]{Peter Awakowicz} 
\author[1]{Martin Muhler*} 
\authormark{Oppotsch \textsc{et al}}

\address[1]{\orgdiv{Laboratory of Industrial Chemistry (LTC)}, \orgname{Ruhr University Bochum}, \orgaddress{\state{Bochum}, \country{Germany}}}
\address[2]{\orgdiv{Chair of Applied Electrodynamics and Plasma Technology (AEPT)}, \orgname{Ruhr University Bochum}, \orgaddress{\state{Bochum}, \country{Germany}}}

\corres{*Martin Muhler, Laboratory of Industrial Chemistry (LTC), Ruhr University Bochum, Universitätsstrasse 150, 44801 Bochum, Germany. \email{martin.muhler@rub.de}}

\abstract[Abstract]{A twin surface dielectric barrier discharge on the microsecond scale was used in combination with a two-component coating to oxidize \SI{300}{ppm} $\nbu$ as a model volatile organic compound to \ch{CO2} and \ch{H2O} in synthetic air at room temperature and at \ele.
The integration of \ch{BaTiO3} as a base material for the coating allowed the successful use of otherwise discharge-ignition-inhibiting materials such as \ch{MnO2}-CuO applied as a full coating. 
Application of pure \ch{BaTiO3} led to highly porous coatings that do not hinder the discharge ignition and show a negligible influence on $\nbu$ conversion while reducing byproduct formation. Thus, \ch{BaTiO3} was identified as a suitable structure-directing agent in two-component coatings, using 1:1 and 1:2 ratios of \ch{BaTiO3}:catalyst coatings.
Coated electrode configurations were compared to their respective uncoated state to highlight the coating-induced changes.
The two-component coatings strongly increased the \ch{CO2} selectivity, reaching a maximum of \SI{91.6}{\%} at an energy density of \SI{450}{J.L^{-1}} and \ele ~ for the 1:2 ratio, corresponding to an increase of \SI{51.0}{\%} relative to the uncoated electrode.
}

\keywords{catalysis, coatings, dielectric barrier discharges, barium titanate, manganese dioxide, volatile organic compounds}

\jnlcitation{\cname{%
\author{Oppotsch T.}, 
\author{Oberste-Beulmann C.},
\author{Böddecker A.},
\author{Hübner G.},
\author{Korolov I.},
\author{Awakowicz P.}, and
\author{Muhler M.}
(\cyear{2025}), 
\ctitle{Overcoming Discharge Inhibition in \textit{n}-Butane Oxidation: Two-Component \ch{BaTiO3} and Mn-Cu Mixed Oxide Coatings}, \cjournal{Plasma Process Polym.}, \cvol{tbd}.}}

\maketitle

\section{Introduction}
   \label{sec:einleitung}

\begin{figure}[h]
                              \centering\includegraphics[width=1.02\linewidth]{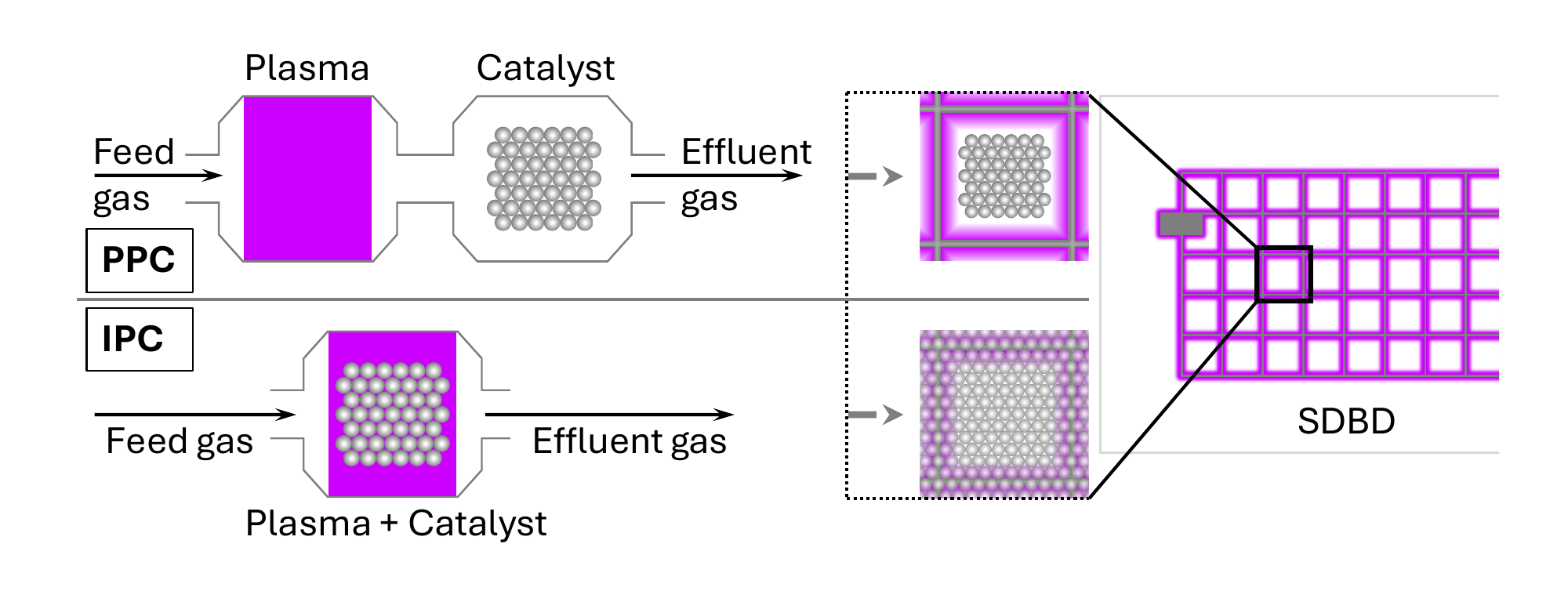}
                              \caption{Schematic illustration of post-plasma catalysis (PPC) and in-plasma catalysis (IPC) and the application of these spatial arrangements to the used SDBD configuration. Adapted from~\cite{Niklas,Kevin,Whitehead}.}
                              \label{fig:ipcppc}
                              \end{figure}

Protection of the environment and human health from detrimental compounds is one of the major goals of the 21\textsuperscript{st} century.
Hence, environmental catalysis is a topic of growing importance\cite{EnvCat1,EnvCat2}.
A large fraction of noxious compounds belongs to the group of volatile organic compounds (VOCs), which are released in various industrial processes and in households, and can lead to an increased risk for several diseases, including hypertension, respiratory diseases and cancer\cite{VOC1,VOC2,VOC3,VOC4,VOC5,VOC6,VOC7}.
Conventional thermal VOC abatement techniques are operated at high temperatures, which lead to high degrees of conversion and a more pronounced selectivity to \ch{CO2}, but are highly energy intensive\cite{thermal}. 
Using catalysts to lower the energy demand simultaneously introduces the risk of catalyst poisoning. 
A pre-treatment of the gas stream can avoid poisoning of the applied catalysts, otherwise supplementary regeneration techniques may be used, or a frequent replacement of the catalyst may be necessary.
Furthermore, the efficiency of thermal VOC abatement is limited for low or spatially or temporally fluctuating VOC loads\cite{Offerhaus}.
Especially in times of high energy costs, efficient plasma-driven processes are further gaining attraction.
Here, non-thermal atmospheric pressure plasmas (NTP) excel due to their high process flexibility and energy efficiency, allowing them to also remove traces of VOCs from fluctuating gas mixtures without any ramp-up time\cite{Bogaerts}. 
When operated in air, mean electron energies in the range of several eV generate reactive oxygen and nitrogen species, such as \ch{O2}(a\textsuperscript{1}\textDelta\textsubscript{g}), O(\textsuperscript{3}P), OH, \ch{O3}, excited nitrogen species, N(\textsuperscript{2}D), N(\textsuperscript{2}P), \ch{N2}(A\textsuperscript{3}\textSigma\SPSB{+}{u}), and nitrogen oxides N\textsubscript{x}O\textsubscript{y} that can interact with the VOC or intermediate reaction products leading to decomposition\cite{Kogelschatz,Lars2022}.
The plasma generation at atmospheric pressure can be achieved by a variety of different types and configurations, mainly by corona discharges or dielectric barrier discharges (DBD), the latter either in a volumetric (VDBD) or surface configuration (SDBD).
DBDs generally consist of two electrodes that are separated by at least one dielectric to prevent the formation of a thermal arc.
The discharge source can be combined with a catalytic material, which can be achieved in an one-stage approach corresponding to in-plasma catalysis (IPC) and a two-stage approach called post-plasma catalysis (PPC).
The IPC configuration is to be considered more beneficial due to potentially strong interactions between plasma and catalyst, but also leads to a more complex system. 
While IPC-VDBD setups, often consisting of a packed catalyst bed exposed to plasma, may generally ensure evenly distributed discharges along the catalyst bed, allowing good contact of plasma and catalyst\cite{PBreactors}, packed-bed plasma reactors are limited by a poor diagnostic access to plasma parameters and by high flow resistances\cite{Kevin,AlexFlow}.
In contrast, the main strengths of an SDBD are its low flow resistance and good optical accessibility. 
Another advantage of the SDBD electrode configuration is the simple scale-up, as demonstrated by Böddecker \textit{\textit{et al.}}\cite{Alex2022,Alex2025}.
Despite its small discharge volume, ionic winds induced by the SDBD generate vortices, which drastically enhance the mixing with the neutral gas domain, thus allowing for reaching degrees of conversion of up to \SI{100}{\%}\cite{AlexFlow,Alex2025,FlowEP2,Offerhaus}. 
\newline
To combine the SDBD with catalytic materials, they are applied in the form of a coating covering the large sides of the electrode configuration. 
The resulting system is used for the VOC removal, exemplarily demonstrated for the oxidation of $\nbu$ (\ch{C4H10}) in synthetic air.
\textit{n}-Butane is a saturated alkane, which is relatively persistent and difficult to oxidize due to the lack of functional groups.
It has been used as a benchmark VOC in a multitude of research works\cite{Niklas,Lars2020,Alex2022,Alex2025}.
Due to the application for environmental purposes, the total oxidation ~ of ~ $\nbu$ ~ to ~ \ch{CO2} ~ according ~ to ~ the ~ reaction \ch{C4H10 + 6.5 O2 -> 4 CO2 + 5 H2O} is required.
The formation of CO, although a more valuable compound for the chemical industry, has to be suppressed in this application due to its toxicity.
A suitable oxidation catalyst commonly used in thermal catalysis is \textalpha-\ch{MnO2}, that has already been implemented in previous work, in which it was not possible to use full coatings, because \textalpha-\ch{MnO2} showed a strong ignition-inhibiting effect\cite{Niklas,NiklasDiss}.
Therefore, a distance of typically \SI{1}{mm} between the coating and the metallic grid had to be left uncoated.
The two resulting configurations ``full coating'' and ``gap coating'' can be directly linked to IPC and PPC, as shown in Figure\,\ref{fig:ipcppc}.
\newline
A major motivation of this work was to surpass the negative effects of catalyst coatings on SDBDs, facilitating novel coating materials and geometries that have not been used before.
This was achieved by introducing a two-component coating using \ch{BaTiO3} (BTO) particles as a basis that acts as a structuring agent for the coating and dilutes the $\mo$ catalyst particles consisting of \ch{MnO2} and CuO (2:1 ratio). 
First, BTO coatings with two different theoretical loadings were used to assess the impact of this first component on the overall process parameters $\nbu$ conversion, CO and \ch{CO2} selectivities, carbon balance, and peak-to-peak voltages required to reach specified values of dissipated power.
Second, the two-component coatings were applied to study the additional influence of the catalyst on these parameters.

\section{Experimental Section}
   \label{sec:expsec}

\subsection{Setup}
   \label{sec:setup} 
\begin{figure}[h]
                              \centering\begin{subfigure}[t]{0.49\linewidth} 
                              \includegraphics[width=.963\linewidth]{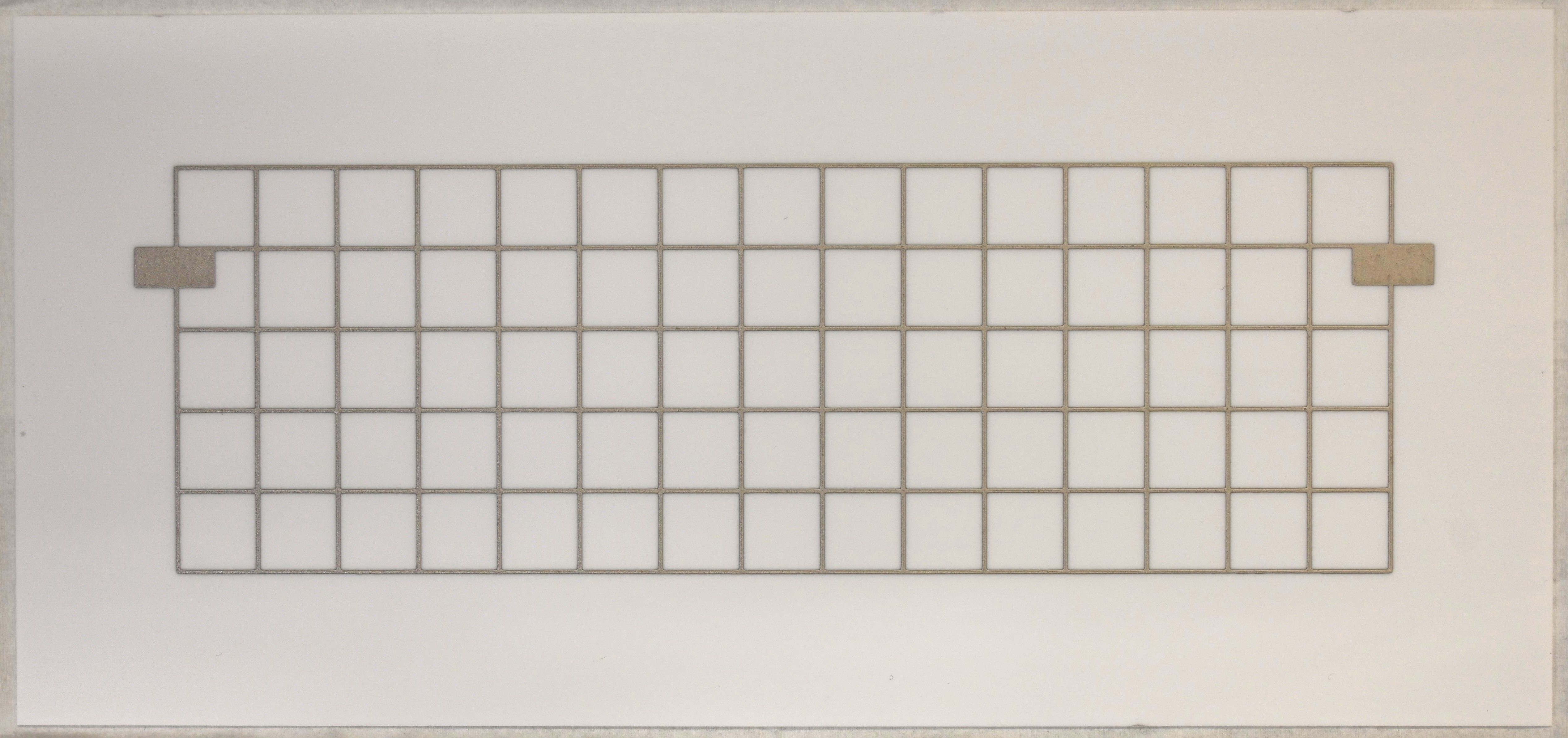}
                                      \pos{\tif{(a)}}
                                   \label{fig:electrode}
                              \end{subfigure}
                              \centering\begin{subfigure}[t]{0.49\linewidth} 
                              \includegraphics[width=1\linewidth]{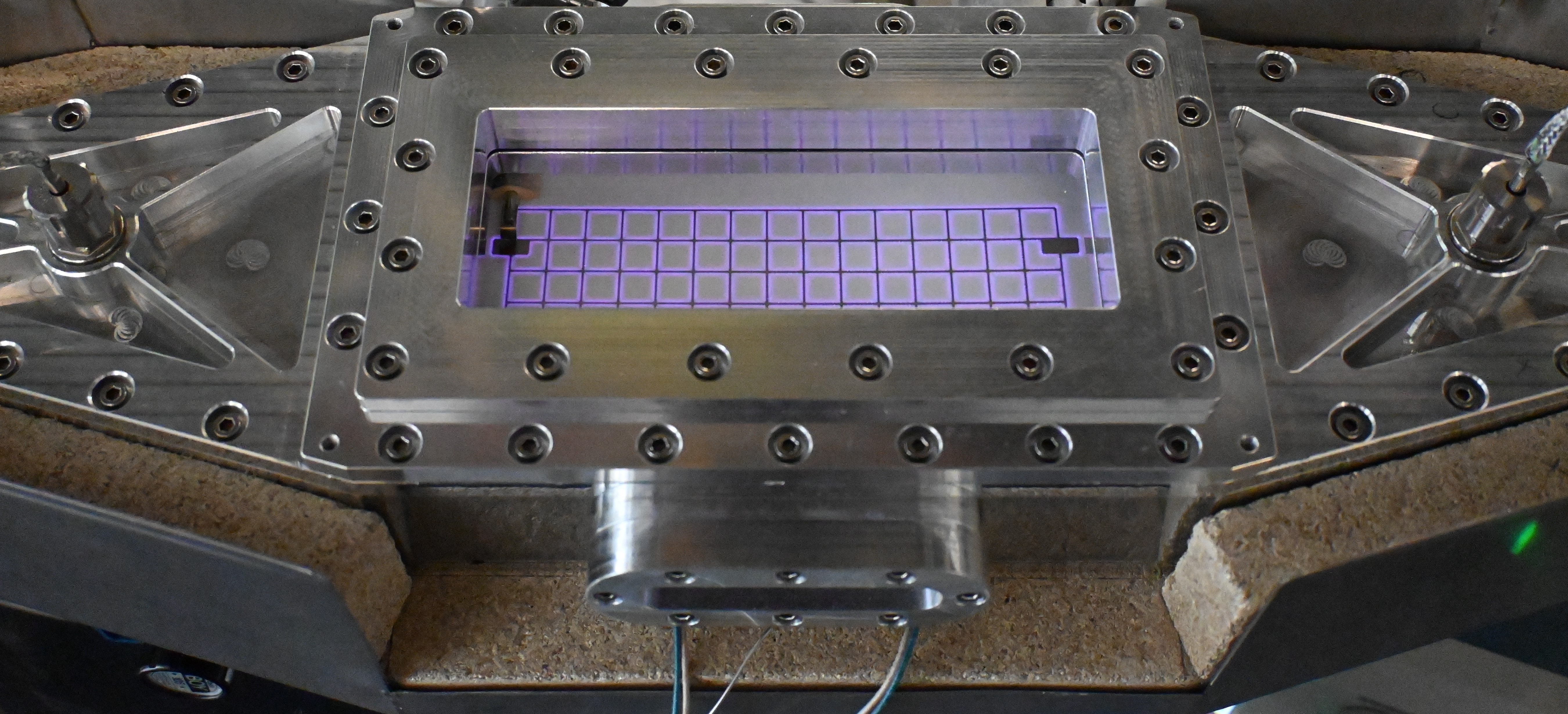}
                                      \pos{\tif{(b)}}
                                   \label{fig:reactor}
                              \end{subfigure}                              
                              \caption{Photographs of the twin SDBD electrode configuration (a) and the used reactor (b) showing an SDBD in operation. The gas flow direction is from left to right. The reactor is placed on a heating plate and thermally insulated. The top window made of quartz glass was used for optical access to the ignition behavior and was replaced by an aluminum lid for the actual VOC oxidation.}
                              \label{fig:electrodereactor}
                              \end{figure}

The discharge source used in this work was a twin SDBD electrode configuration (Alumina Systems GmbH, Germany) already described and used in previous works\cite{Niklas,Offerhaus,Lars2020,Lars2022,RTNS,Alex2022,Alex2025}.
The electrode configuration consists of an \textalpha-\ch{Al2O3} plate spanning \qtyproduct{190 x 88 x 0.635}{mm}, which serves as a dielectric barrier to separate two chemically nickelled molybdenum and manganese silicate (ratio 80/20) metal grids that were symmetrically screen‐printed on its two large sides.
The grid structure comprised \qtyproduct{5x15}{} \SI{100}{mm^2} large squares with a grid line width of \SI{.45}{mm}.
A photograph of the electrode configuration is shown in Figure\,\ref{fig:electrodereactor}a. 

The voltage supply (G2000, Redline Technologies, Germany) generated square unipolar pulses (\qtyrange[range-units = single, range-phrase= --]{0}{300}{V}) that were applied to the electrode with a pulse repetition frequency of \SI{4}{kHz}.
High voltages were obtained by transforming the output voltages from the generator.
The inductance of the secondary side of the transformer together with the capacitance of the electrode configuration formed a series resonance circuit with an eigenfrequency of about \SI{86}{kHz}.
Energy dissipation to the system led to damping of the resulting sine waves.
A current probe (bandwidth up to 300 MHz; Model 6585; Pearson Electronics) and a high-voltage probe (P6015A; Tektronix) were used, which were connected to a digital oscilloscope (DPO5204B; Tektronix) to record these quantities over time.
More detailed information regarding the electrical setup can be found in previous work by Schücke \textit{et al.}\cite{Lars2020,LarsDiss}, Nguyen-Smith \textit{et al.}\cite{RTNS}, and Böddecker \textit{et al.}\cite{Alex2022,Alex2025}.
According to Equation\,\ref{eq:Pdiss}, current and voltage were used to calculate the dissipated power in real time, which was the main set parameter for the VOC oxidation measurement campaign.
$P_{\mathrm{diss}}$ was varied in the range of \qtyrange[range-units = single, range-phrase= --]{15}{75}{W} as further described in section\,\ref{sec:procedure}.

							\begin{equation}
							P_{\mathrm{diss}} = f_{\mathrm{rep}} \cdot \int_{0}^{T_{\mathrm{p}}}  U(t) \cdot I(t)  \; \mathrm{d}t.
							\label{eq:Pdiss}
							\end{equation}

Here, $f_{\mathrm{rep}}$ is the repetition frequency, $T_{\mathrm{p}}$ denotes the duration of the pulse, $U(t)$ and $I(t)$ are the measured time-resolved voltage and electrical current, respectively.  
The displacement current $C \dv{U(t)}{t}$ with capacitance $C$ can be safely disregarded for integration over a full cycle and was therefore not included in Equation\,\ref{eq:Pdiss}. 
The main part of the setup was the externally heatable reactor, which had also been used in former works\cite{Niklas}.
The reactor was made of aluminum with the outer dimensions of \qtyproduct{540 x 170 x 51}{mm}.
An electrode configuration was placed in the \qtyproduct{105 x 19}{mm} large rectangular plane and held by polyether ether ketone mountings right in the center of the reactor.
A photograph of the reactor is shown in Figure\,\ref{fig:electrodereactor}b. 
The shown quartz glass window enabled optical accessibility, whereas a solid aluminum lid was used for the actual VOC oxidation measurements.
The reactor was placed on a custom-tailored heating plate driven by seven heating cartridges and thermally insulated.
To account for Joule heating, two cartridges were placed underneath the reactor inlet and five directly under the center, resulting in an asymmetric arrangement. 
The bulk gas temperature inside the reactor was monitored by two thermocouples that were placed at a distance of \SI{85}{mm} from the electrode configuration in the inlet and outlet direction. 
Besides the main heating, the feed gas was preheated in a copper block that surrounded the tubing, which was also heated by cartridges but 40~\textcelsius ~ higher than the set temperature of \ele.
Preheating was required due to the relatively high volumetric flow rate of \SI{10}{sL.min^{-1}} (standard liter per minute, slm) and insulation losses. 
The feed gas consisted of \ch{N2} (\SI{99.999}{\%}), \ch{O2} (\SI{99.998}{\%}), and $\nbu$ (\SI{99.5}{\%}) supplied by Air Liquide. 
Mass flow controllers (MFC, Bronkhorst High-Tech B.V., Germany) were used to adjust the composition to synthetic air comprising \SI{79.5}{\%} \ch{N2}, \SI{20.5}{\%} \ch{O2}, admixed with \SI{300}{ppm} of $\nbu$.
Reactants and products were monitored by an online multichannel analyzer (MCA, X‐stream XEGP, Emerson Process Management, Germany) equipped with non-dispersive infrared sensors for CO and $\COt$ (\qtyrange[range-units = single, range-phrase= --]{0}{10000}{ppm} each), and $\nbu$ (\qtyrange[range-units = single, range-phrase= --]{0}{500}{ppm}), and a paramagnetic \ch{O2} sensor (\qtyrange[range-units = single, range-phrase= --]{0}{25}{\%}).

\subsection{Conversion Measurements}
   \label{sec:procedure}  

Conversion measurements were generally performed at room temperature and at \ele. 
At \ele, the catalyst is known to be active in CO oxidation to \ch{CO2}, but cannot oxidize $\nbu$ \cite{Niklas,NiklasDiss,KevinCOtherm}.
In both cases, Joule heating was not further considered as it had only a minor contribution. 
A measurement sequence comprised the collection of effluent $\nbu$, CO, and \ch{CO2} mole fractions and applied peak-to-peak voltages for a dissipated power range of \qtyrange[range-units = single, range-phrase= --]{15}{65}{W} and \qtyrange[range-units = single, range-phrase= --]{15}{75}{W} at room temperature and at \ele ~ in \SI{10}{W} increments.
The measurement without discharge ignition was denoted \SI{0}{W} and served as an individual reference for each sequence corresponding to initial values and was performed before and after each sequence.
Each set point was held for \SI{1000}{s} to ensure steady-state conditions at the respective dissipated power and the last \SI{250}{s} were time-averaged to determine mole fractions and peak-to-peak voltage.
Due to intrinsic differences between individual electrode configurations, a full set of measurements was first performed for any electrode configuration to provide reference data and then repeated after the coating step.  
To account for statistical deviations, three measurements were performed for each of the BTO series and then averaged. 
For the second part involving the two-component coatings, only single measurements were performed.
Finally, before first use and after coating, an activation measurement was performed in which the electrode configuration was exposed to synthetic air and the discharge was ignited at a dissipated power of \SI{45}{W}.
This was required because of an aging effect of the electrode configuration when exposed to operating conditions leading to partial erosion of the metallic grid\cite{RTNS}.
The second and shorter activation served the purpose of removing potential residuals from the coating.
Conversion ($X_{\nbu}$) is determined according to Equation\,\ref{eq:X}, while Equation\,\ref{eq:S} refers to selectivity ($S_{\COx}$) to either CO or \ch{CO2}, expressed as $\COx$.
The carbon balance ($CB$) determined according to Equation\,\ref{eq:CB} is another important parameter to prove whether all carbon-containing compounds were detected.

							\begin{equation}
							X_{\nbu}=\frac{y_{\nbui}-y_{\nbu}}{y_{\nbui}} 
							\label{eq:X}
							\end{equation}

Here, $y_{\nbui}$ and $y_{\nbu}$ are the mole fractions of \textit{n}-butane in the feed and in the effluent, respectively.
In Equation\,\ref{eq:S}, the stoichiometric factors $\nu_{\nbu}$ and $\nu_{\COx}$ for \textit{n}-butane and CO/\ch{CO2} were used, respectively.
$CB$ counts and relates the number of carbon atoms that enter and leave the reactor considering $\nbu$, CO, and $\COt$.
Values around \SI{100}{\%} are ideal but due to experimental limitations and statistical fluctuations, the tolerable regime is \SI{90}{\%}--\SI{100}{\%}. 

							\begin{equation}\begin{split}
							S_{\COx} & = \frac{y_{\COx}-y_{\COxi}} {y_{\nbui}-y_{\nbu}} \cdot \frac{\nu_{\nbu}}{\nu_{\COx}} \\ 
						      & = \frac{y_{\COx}-y_{\COxi}} {y_{\nbui}-y_{\nbu}} \cdot \frac{1}{4}
							\end{split}	
							\label{eq:S}
							\end{equation}

							\begin{equation}
							CB = \frac{y_{\CO}-y_{\COi}+y_{\COt}-y_{\COti}+4 \cdot y_{\nbu}} {4 \cdot y_{\nbui}}  	
							\label{eq:CB}
							\end{equation}

\subsection{Coatings}
   \label{sec:coatings}
Coatings were applied on both sides of the electrode configuration using commercially available \ch{BaTiO3} (cubic nano powder, \SI{50}{nm}, \SI{99.9}{\%}, Sigma-Aldrich) and $\mo$ (Molecular Products Group), which is a 2:1 combination of \ch{MnO2} and CuO and referred to as the catalyst.
Prior to use, the catalyst was ground, pressed and sieved to obtain a fraction with a particle size of \SI{<250}{\textmu m}.
Although consisting of two components, $\mo$ is considered a single component here with the main focus on \ch{MnO2}. 
Coatings comprising both BTO and catalyst were obtained by mixing the two materials using equimolar amounts (1:1) and the doubled molar amount (1:2) of $\mo$. \\
The general coating procedure was similar to the one described previously in our former work, for which the same custom-tailored spray coater was used\cite{Niklas}.
One major difference was the preparation of full coatings in a direct approach without the use of a mask to prevent the metal grid from being coated.
An \textit{iso}-propanol suspension containing \SI{1.2}{mg.ml^{-3}} of the material to be applied as a coating was pumped to a nozzle, atomized by compressed air and thus sprayed onto the surface of the electrode configuration. 
The target was placed on a heating plate and heated to 225\,\textcelsius ~to vaporize the solvent rapidly on contact.
The nozzle was then moved by motors in two dimensions in \SI{2}{mm} increments following a predefined program with an overspray on the sides to ensure a homogeneous coating. 
Afterward, the coating homogeneity was controlled with a laser scanning microscope (LSM; VK9710; Keyence), which can be operated either in an optical or laser mode.

\section{Results}
   \label{sec:results}

\subsection{Coating structure and ignition behavior}
   \label{sec:coating_ignition}

\begin{figure}[h]
                              \centering\begin{subfigure}[t]{0.32\linewidth}
                              \centering
                              \includegraphics[width=\textwidth]{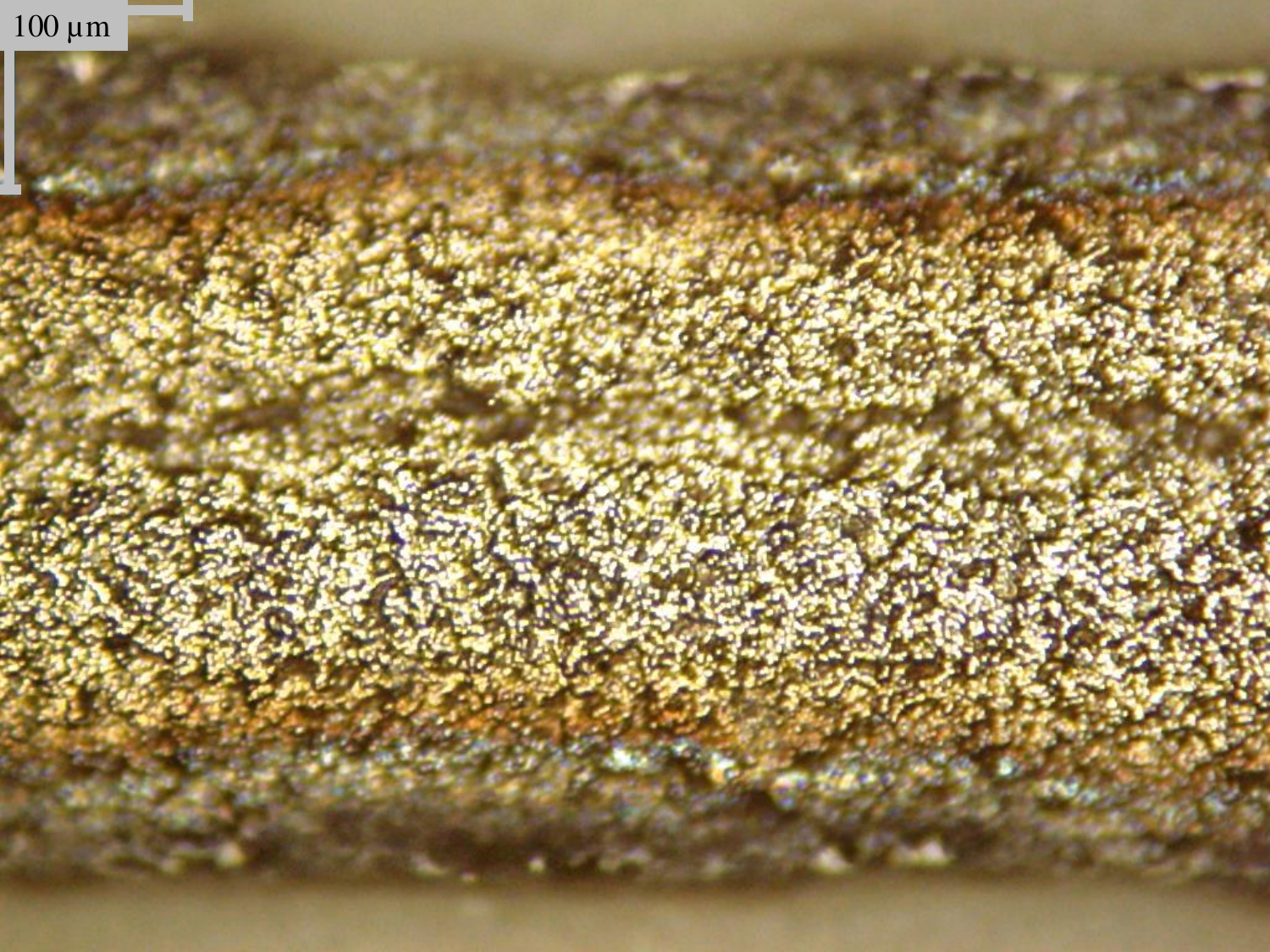}
                                      \pos{\tif{(a)}}
                                   \label{fig:Mikroskopie_a}
                              \end{subfigure}\hfill     
                              \centering\begin{subfigure}[t]{0.32\linewidth}
                              \centering
                              \includegraphics[width=\linewidth]{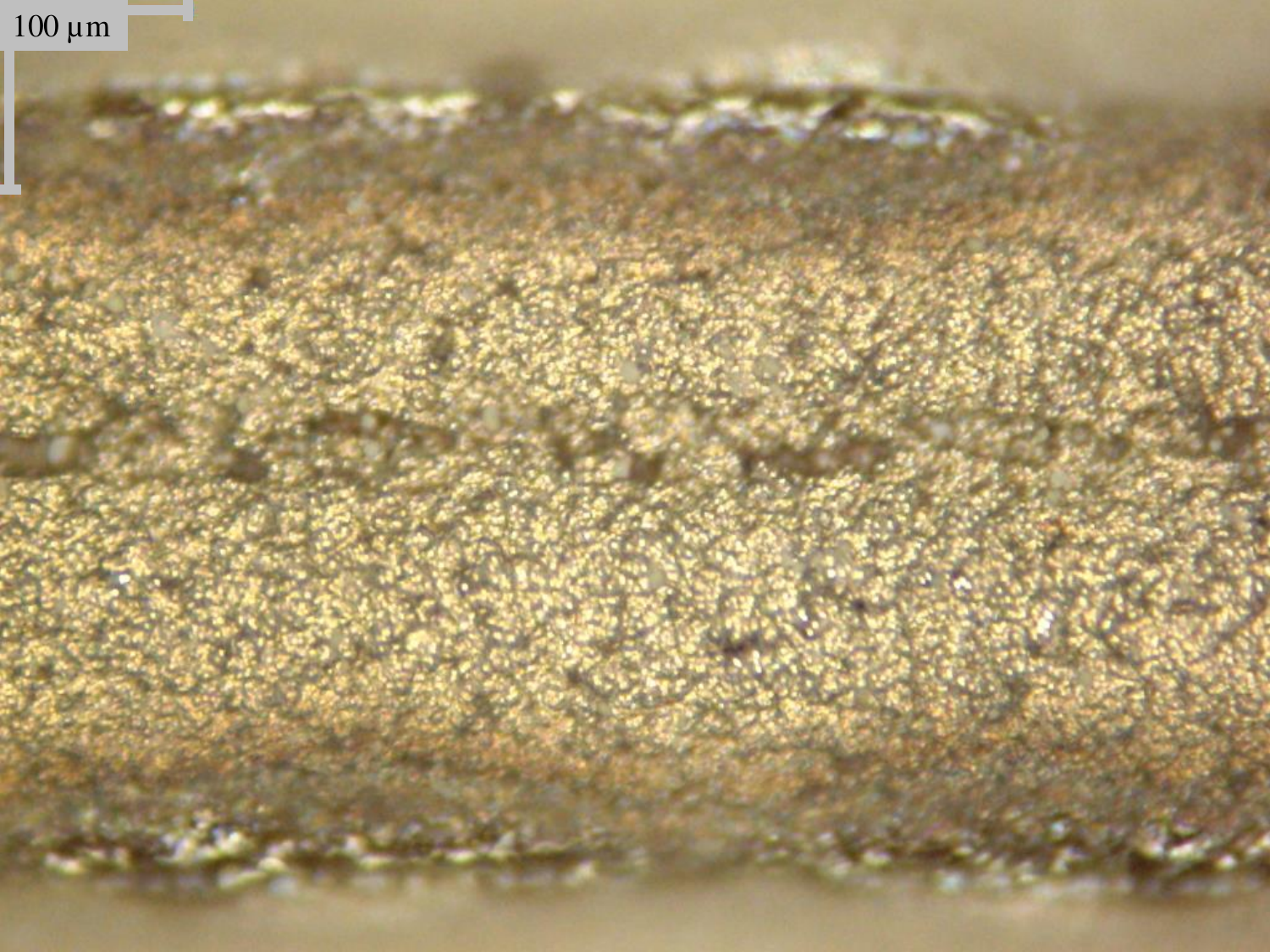}
                                      \pos{\tif{(b)}}
                                   \label{fig:Mikroskopie_b}
                              \end{subfigure}\hfill
                              \centering\begin{subfigure}[t]{0.32\linewidth}
                              \centering
                              \includegraphics[width=\linewidth]{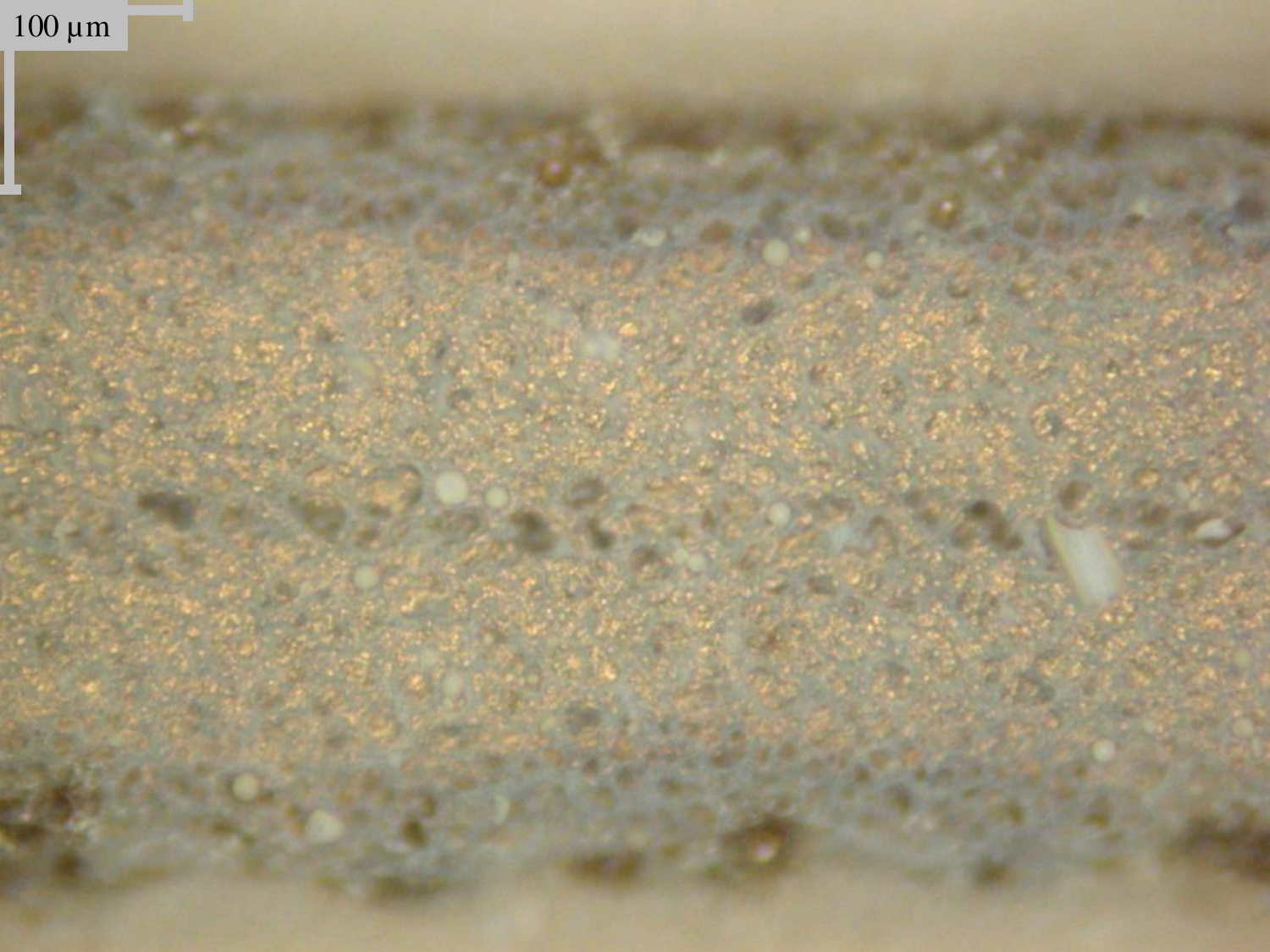}
                                      \pos{\tif{(c)}}
                                   \label{fig:Mikroskopie_c}
                              \end{subfigure}
                              \caption{Optical microscopy images of a single grid line at 20x magnification. From left to right: uncoated (a) and BTO-coated electrode configurations with loadings of \SI{.3}{mg.cm^{-2}} (b) and \SI{3}{mg.cm^{-2}} (c).}
                              \label{fig:Mikroskopie}
                              \end{figure}

The basis for achieving a catalyst-containing coating was to use a material that shows no discharge ignition-inhibiting properties at all.
As test measurements with BTO showed promising results, a full coating with BTO was prepared instead of the previously applied mask-based coating with gaps between the coating and the metallic grid.
Two different theoretical BTO loadings on the electrode configurations of \SI{.3}{mg.cm^{-2}} and \SI{3}{mg.cm^{-2}} were used, with the latter being the usual value for coating related to the used electrode design. 
Before assessing the actual performance in the oxidation reaction of the model VOC $\nbu$, more fundamental tests were performed.
The presence and uniformity of the BTO coating were visualized by optical microscopy shown in Figure\,\ref{fig:Mikroskopie}.
All three images comprising an uncoated and the two coated electrode configurations represent a single electrode grid line after exposure to a discharge in synthetic air with a magnification factor of 20.
The discharge exposure left visible marks in the form of discoloration and partial erosion of the electrode grid.
This erosion has been investigated and discussed by Ngyuen-Smith \textit{et al.}\cite{RTNS}.
The middle trench structure was already present on unused electrode configurations, originating from the production process.
Although the three-dimensional structure of the grid and its reflective nature prevent a simultaneous high-resolution image of the whole grid line, the latter effect can be exploited to assess the coating success: in addition to the small spherical off-white deposits, the image representing the lower BTO loading appears matte and blurred compared with the uncoated grid.
This effect can be explained by BTO nanoparticle deposition on the grid line.
With increasing loading, the deposited BTO is more clearly visible, forming a highly porous, net-like structure that was found to spread uniformly across the entire electrode grid. 
Irregularly, additional BTO spheres fill the voids of the main structure. 

\begin{figure}[h]
                              \centering\begin{subfigure}[t]{0.49\linewidth} 
                              \includegraphics[width=.992\linewidth]{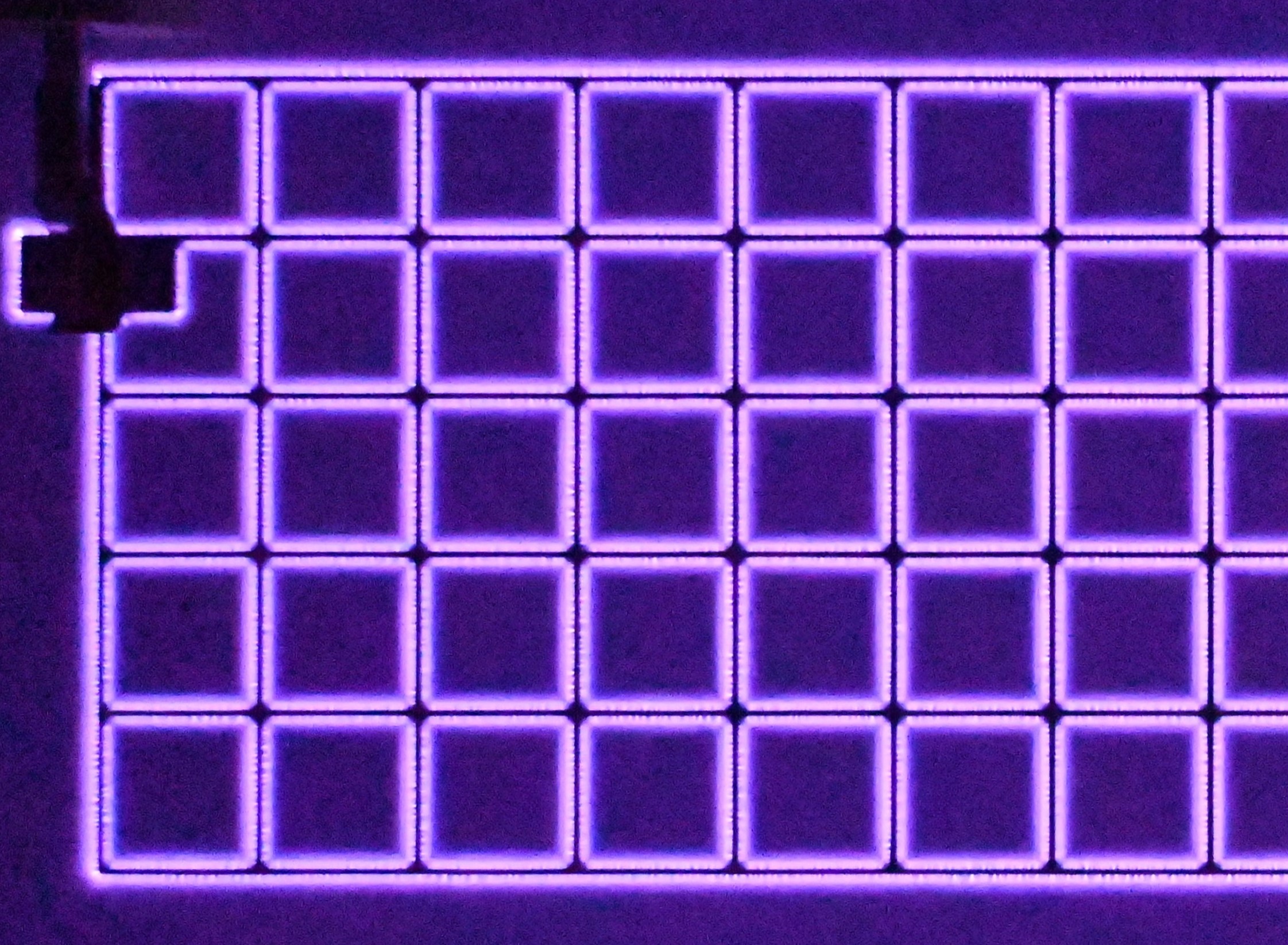}
                                      \pos{\tif{(a)}}
                                   \label{fig:ignition_a}
                              \end{subfigure}
                              \centering\begin{subfigure}[t]{0.49\linewidth} 
                              \includegraphics[width=1\linewidth]{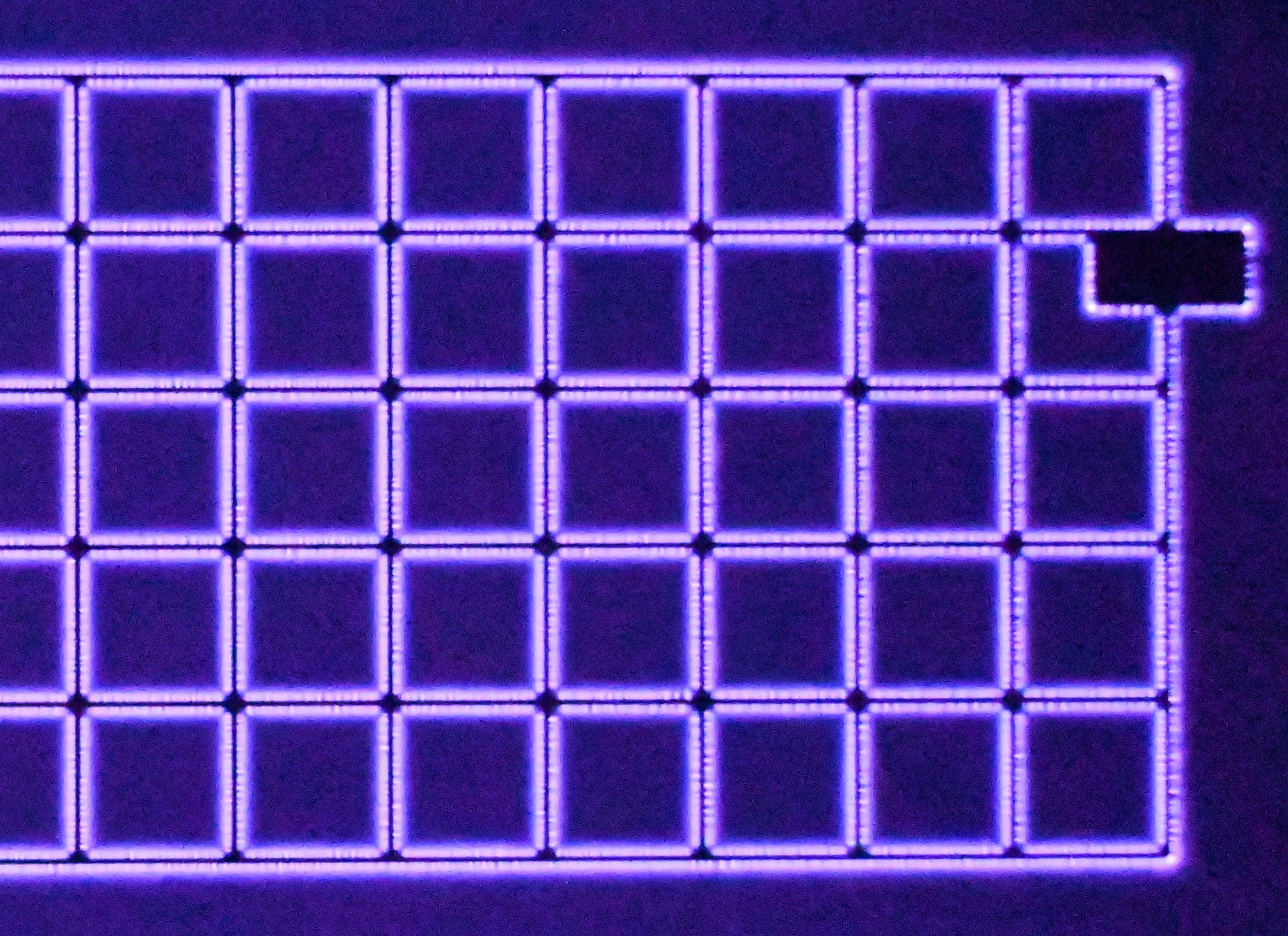}
                                      \put(-115,10){\tif{(b)}}
                                   \label{fig:ignition_b}
                              \end{subfigure}                              
                              \caption{Photographs of the BTO-coated electrode configurations with loadings of \SI{.3}{mg.cm^{-2}} (a) and \SI{3}{mg.cm^{-2}} (b) ignited at a peak-to-peak voltage of \SI{8.5}{kV} in synthetic air.}
                              \label{fig:ignition}
                              \end{figure}

\begin{figure}[h]
                              \centering
                              \includegraphics[width=\linewidth]{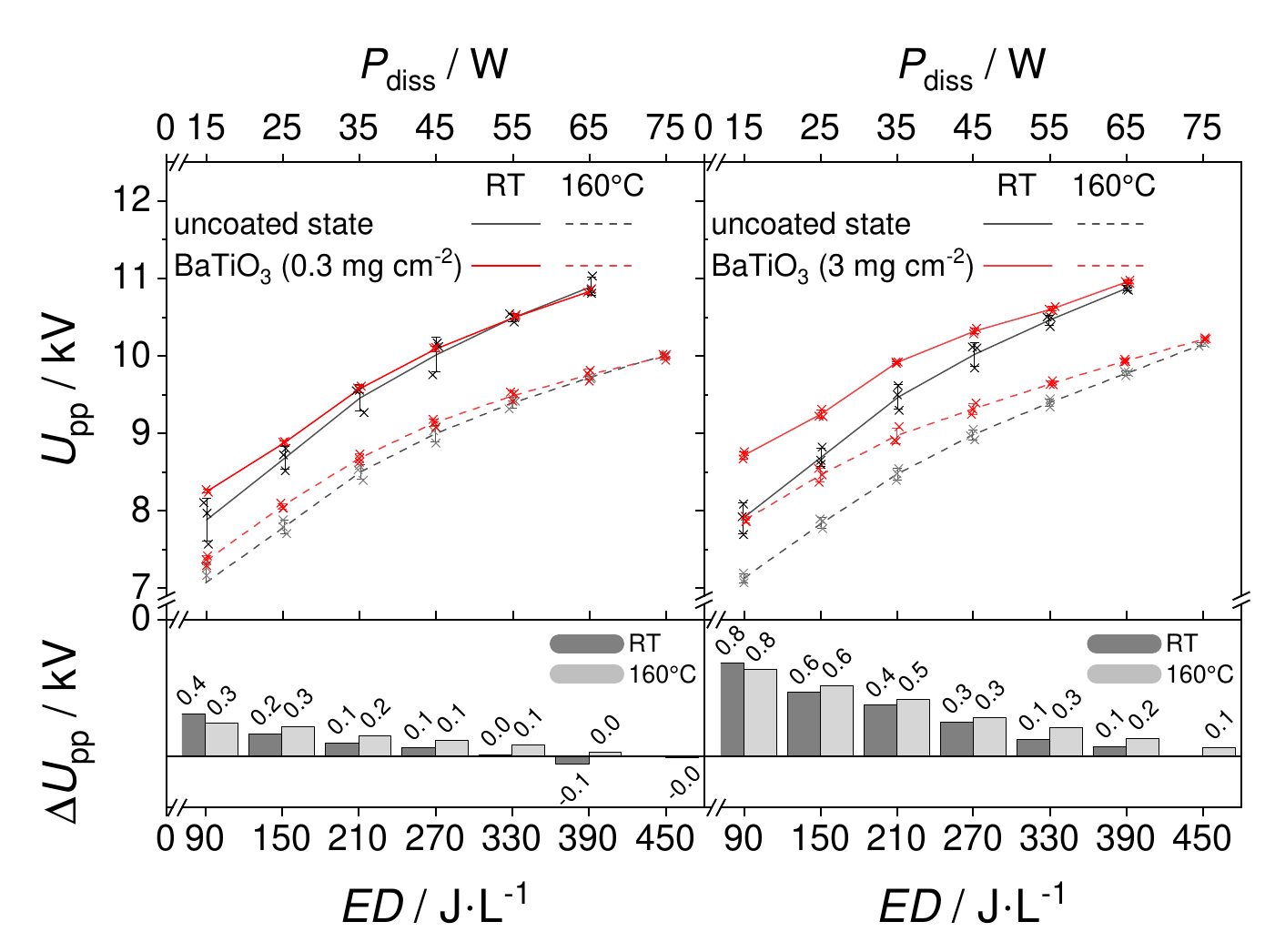}
                              \caption{Applied peak-to-peak voltages required to reach specific values of dissipated power in the range of \SI{15}{W} to \SI{75}{W} for both electrode configurations at room temperature and at \ele ~ for the uncoated and coated state. To highlight coating-induced changes, $\Delta U_{\mathrm{pp}}$, is shown in the lower section.}
                              \label{fig:Upp}
                              \end{figure}

To demonstrate the successful uniform ignition for both coatings, Figure\,\ref{fig:ignition} shows exemplary photographs of the SDBD at a peak-to-peak voltage of \SI{8.5}{kV} ignited in synthetic air.
The emission, mainly determined by the deexcitation of \ch{N2} species, is used as an indicator for the propagation of the streamers. 
Once again, the filamentary nature of the discharge is clearly visible\cite{RTNS} and changes, if at all, only marginally with increasing BTO loading. 
The slightly differing color observed in Figure\,\ref{fig:ignition} between the two half-sided SDBDs can be traced back to different camera settings.
Still, a minor reduction of the light intensity for the higher BTO loading may be coupled to the higher required voltage as observed in Figure\,\ref{fig:Upp}, see the discussion below.
Areas with less intense or even no light emission were not found, indicating a homogeneous BTO coating.
Furthermore, the discharge region was unaffected by an increase in the BTO loading as well. 
Hence, BTO meets the expectations as a non-discharge-inhibiting material, rendering it a promising basis for two-component coatings. 

Despite its absence of ignition-inhibiting properties, the presence of BTO still affects the discharge behavior.
One aspect is the applied voltage required to reach the same power levels as before.
Figure\,\ref{fig:Upp} shows the applied peak-to-peak voltage needed to reach the set values of power in comparison to the respective electrode configuration in its uncoated state.
These dissipated powers (Equation\,\ref{eq:Pdiss}) were \SI{15}{W} to \SI{65}{W} in steps of \SI{10}{W} at room temperature, while in the heated state at \ele, the upper limit was increased to \SI{75}{W}.
The upper limit was determined according to the occurrence and probability of arcing to minimize the risk of partly damaging the setup. 
For a better comparison, the corresponding energy density ($ED$) was calculated according to Equation\,\ref{eq:ED} and included in Figure\,\ref{fig:Upp} as well. 
Since the main focus was on the changes induced by the application of the BTO coating, $\Delta U_{\mathrm{pp}}$ represents the voltage measured for a coated electrode referenced to its own uncoated state. 

							\begin{equation}
							ED=\frac{P_{\mathrm{diss}}}{\dot{V}} 
							\label{eq:ED}
							\end{equation}

With ~ the ~ volumetric ~ gas ~ flow ~ rate ~ $\dot{V}=\mathrm{\SI{10}{slm}}$, ~ Equation\,\ref{eq:ED} ~ can ~ be simplified ~ to ${ED = P_{\mathrm{diss}} \cdot \mathrm{\SI{6}{s.L^{-1}}}}$. 

Two main observations can be extracted from Figure\,\ref{fig:Upp}: first, the presence of a BTO coating increases the necessary voltage to reach the same dissipated power compared with the uncoated state.
For lower dissipated power values, this effect is more pronounced and fades out when reaching higher dissipated power levels.
Second, the higher BTO loading leads to a significantly increased required voltage to reach the same powers at lower values of dissipated power relative to the uncoated state of the electrode configuration.
This increase is roughly double in relation to the lower loading case.
Taking into account that the theoretical loading ratio of the two coatings is ten, a doubling of the value is, however, a much lower change than expected.
Generally, higher values of dissipated power are expected to be correlated with larger quantities of reactive species potentially contributing to chemical reactions.
For these more relevant values, the presence of the coating led to a decrease of the necessary voltage.
Due to intrinsic differences between electrode configurations, manifesting as a lowered peak-to-peak voltage required to reach the target dissipated powers for the second electrode configuration in comparison to the uncoated first one, the absolute $U_{\mathrm{pp}}$ for both BTO loadings are still similar.
Those intrinsic differences cannot be completely avoided but only reduced by the initial activation step, again highlighting the necessity of determining parameters for both the uncoated and coated state of a single electrode configuration.

\subsection{Plasma-assisted oxidation of \textit{n}-butane}
   \label{sec:XS}

\begin{figure}[h]
                              \centering
                              \includegraphics[width=\linewidth]{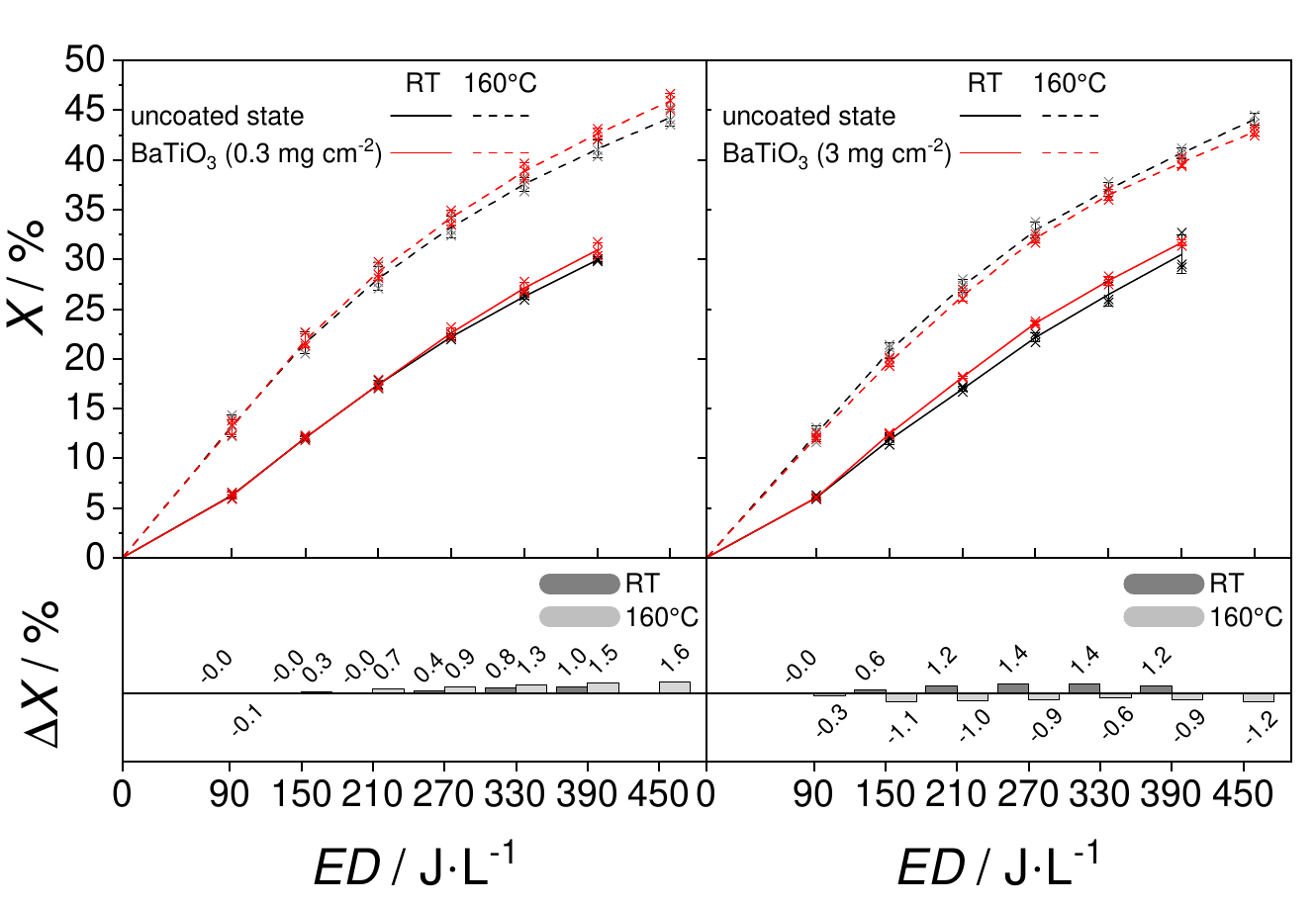}
                              \caption{Conversion for both electrode configurations coated with BTO at room temperature (solid lines) and at \ele (dashed lines) as a function of the energy density. The upper part of the diagram compares the coated (red) and the uncoated (black) states of the respective electrode configuration, while the lower part shows the conversion difference.}
                              \label{fig:BTO_X}
                              \end{figure}

\begin{figure*}[h]
                              \centering\begin{subfigure}[t]{0.66\linewidth} 
                              \includegraphics[width=1\linewidth]{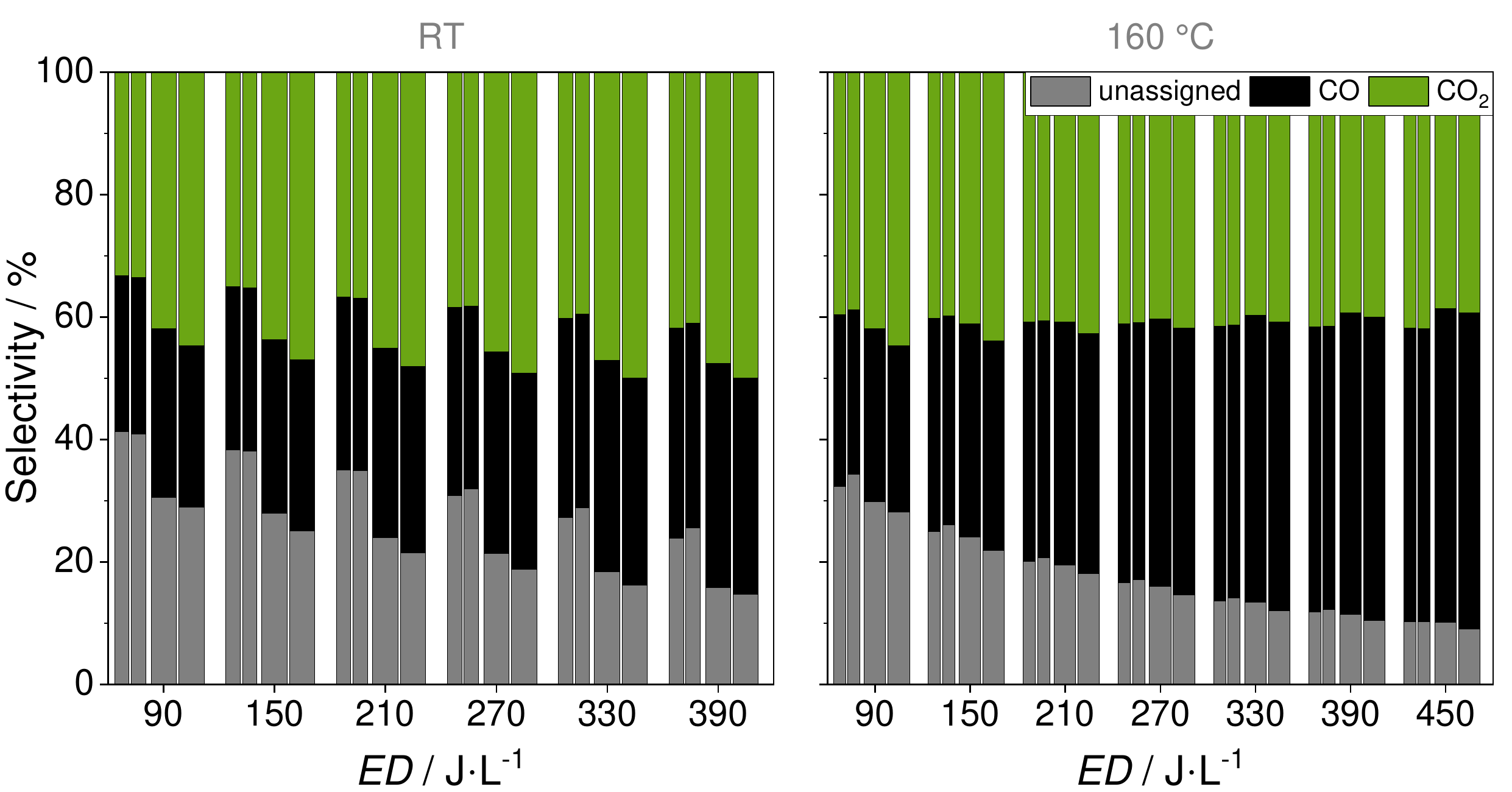}  
                                    \put(-30,30){\tif{(a)}}
                                \label{fig:BTO_S}
                              \end{subfigure}
                              \hfill
                              \centering\begin{subfigure}[t]{0.3055\linewidth} 
                              \includegraphics[width=1\linewidth]{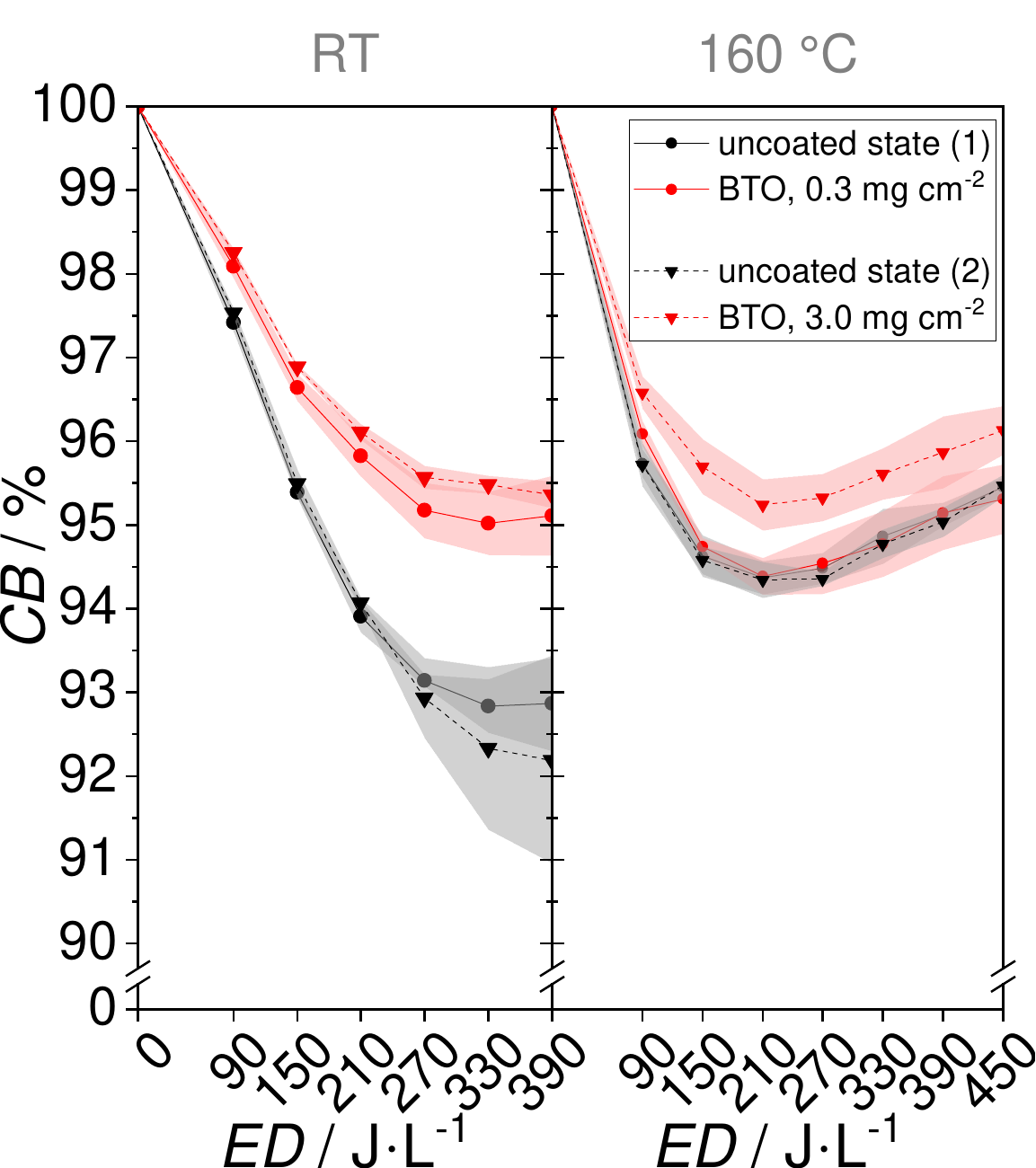}
                                    \put(-30,30){\tif{(b)}}
                                \label{fig:BTO_CB}
                              \end{subfigure}                              
                              \caption{Selectivities to CO, \ch{CO2}, and unidentified products (a) and carbon balance (b) determined according to Equations~\ref{eq:S}~and~\ref{eq:CB} for both electrode configurations coated with BTO at room temperature and \ele. For selectivity data, the first two columns represent the uncoated states of the subsequently coated electrode configurations with the lower and higher loading of \SI{.3}{mg.cm^{-2}} and \SI{3}{mg.cm^{-2}}, respectively, which are then shown as columns three and four in the sequence.}
                              \label{fig:BTO_SCB}
                              \end{figure*}

Both BTO-coated electrode configurations were subsequently used for the plasma-assisted oxidation of \textit{n}-butane to determine the BTO influence on conversion and the selectivities to the main products CO and \ch{CO2}.
Since the overall aim is to remove VOCs for environmental purposes, high conversion and a high selectivity of \ch{CO2} are desired.
Conversion, selectivity, and $CB$ were determined for both the uncoated and the coated state of the two electrode configurations.
Figure\,\ref{fig:BTO_X} shows the achieved $\nbu$ conversion as a function of the energy density, whereas CO/\ch{CO2} selectivities along with the unidentified fraction are summarized in Figure\,\ref{fig:BTO_SCB}a next to the related $CB$ shown in Figure\,\ref{fig:BTO_SCB}b. 
All results shown in Figure\,\ref{fig:BTO_X} and Figure\,\ref{fig:BTO_SCB} represent mean values of three measurements. 
Error bars were obtained by calculating the standard deviation. 

Figure\,\ref{fig:BTO_X} shows the direct unreferenced degrees of conversion in the upper part, whereas the lower part represents the subtraction between the coated and uncoated electrode to further highlight potential changes induced by the coating.
Figure\,\ref{fig:BTO_X} clearly demonstrates that the most dominant changes are induced by heat and dissipated power. 
Heating to \ele ~ results in a nearly constant vertically shifted curve by approximately \SI{10}{\%}.
The expected conversion increase with dissipated power proceeds in a sigmoidal trend at room temperature and in a more logarithmic way at \ele.
However, a final value is not reached in any case and would require higher energy densities. 
The maximum achieved degrees of conversion at room temperature are \SI{29.9}{\%} and \SI{31.0}{\%} for the uncoated and coated state of the lower loading of \SI{0.3}{mg.cm^{-2}}, respectively, which increase to \SI{44.3}{\%} and \SI{45.9}{\%} at \ele ~ for the highest dissipated power.
For the higher BTO loading of \SI{3}{mg.cm^{-2}}, \SI{30.5}{\%} and \SI{31.7}{\%} are reached at room temperature for the two states and \SI{44.1}{\%} and \SI{42.8}{\%} at \ele.
Contrasting these two important parameters, both the presence of the BTO coating and its loading show a rather negligible influence on conversion.
The lower loading of \SI{.3}{mg.cm^{-2}} generally leads to slight increases in conversion, whereas the higher loading of \SI{3}{mg.cm^{-2}} leads to an increase at room temperature and a decrease when heated.
Overall, the coating-induced conversion changes make up less than \SI{2}{\%} in any case and are therefore considered negligible. 
\\ Selectivities to CO and \ch{CO2} and all unidentified products are summarized in Figure\,\ref{fig:BTO_SCB}a. 
The two thinner columns in the sequence represent the values obtained with the uncoated states of the subsequently coated electrode configurations with the lower and the higher BTO loading. 
Selectivities for the coated states are summarized by the thicker columns.
The left part of Figure\,\ref{fig:BTO_SCB}a shows selectivities at room temperature, whereas the right side shows the measurements at \ele. 
The $CB$ is shown in Figure\,\ref{fig:BTO_SCB}b. 
Coated states are shown in red, whereas black refers to the uncoated states.
Errors are indicated by a shaded area.
Differing from other plots in this work, solid and dashed lines here distinguish between the two loadings to fit the depiction of the selectivity diagram, where the two temperature regimes are split.  

CO and \ch{CO2} are the two expected main products.
Nonetheless, Figure\,\ref{fig:BTO_SCB}a shows that a rather large fraction of products is not covered by these two main compounds. 
The share of unidentified products, determined by subtracting the sum of CO and $\COt$ selectivity from \SI{100}{\%} ranges from \SI{9.0}{\%} to \SI{41.3}{\%}.
Due to the lower conversion degrees, especially for lower dissipated power, these values present only a rough estimation and suffer from error propagation. 
Still, the results clearly indicate that there must be additional products, and Figure\,\ref{fig:BTO_SCB}b supports this finding, since the $CB$ is below 100\%. 
Especially for lower degrees of conversion, unreacted $\nbu$ contributes significantly to the $CB$, so the large fraction of unassigned products is negligible in terms of the $CB$.
Both the coated states (shown in red) of the two electrode configurations at room temperature and elevated temperature and the uncoated states (shown in black) at elevated temperature settle at around \SI{95}{\%} of the $CB$, and only the uncoated states for the two electrodes at room temperature exhibit lower $CB$ shares. 
Most likely, the fraction of unassigned compounds comprises intermediates in the CO/\ch{CO2} formation, such as formaldehyde or different alcohols, aldehydes, ketones, or acids, as determined by Schücke \cite{LarsDiss}.
Due to the reactive nature of an SDBD, a large variety of compounds could be considered, but a final assignment is complicated by overlapping mass-to-charge ($m/z$) signals when applying mass spectrometry.
Especially at \ele, the share of unassigned products diminishes with increasing dissipated power, most likely due to the coupled increase in $\nbu$ conversion reducing deviations but also due to more harsh reaction conditions. 
At room temperature, the change occurs mainly in favor of $\COt$, and of CO at \ele. 
Departing from its lowest value at \SI{90}{J.L^{-1}}, the selectivity to CO increases in the same way as conversion and independent of the loading, whose presence leads to an increase of up to \SI{3.8}{\%}.
Selectivity to \ch{CO2} exhibits a temperature-dependent trend inversion. 
At room temperature, the BTO coating leads to a significant \ch{CO2} selectivity enhancement, which is more pronounced at lower energy densities (with \SI{8.8}{\%} and \SI{11.1}{\%} at \SI{90}{J.L^{-1}} compared with \SI{5.8}{\%} and \SI{9.0}{\%} at \SI{390}{J.L^{-1}} for the lower and the higher loading, respectively).
Repeating the measurements with the coating present on the electrode configuration leads to an initially higher \ch{CO2} selectivity.
In contrast to the uncoated electrode configuration at room temperature, heating leads to a decreasing \ch{CO2} selectivity for both loadings, which even falls below the uncoated values.
The lower loading leads to an initial increase of \SI{2.2}{\%} at \SI{90}{J.L^{-1}} but diminishes to a decrease of \SI{3.1}{\%} at \SI{390}{J.L^{-1}}, whereas the higher loading starts with a plus of \SI{5.8}{\%} at \SI{90}{J.L^{-1}}, which reduces to a minus of \SI{2.5}{\%} at \SI{390}{J.L^{-1}}.
A lower \ch{CO2} selectivity can either originate from its possibly reduced formation or from an initialized or more pronounced \ch{CO2} splitting to CO. 
Overall, changes induced by the BTO coatings mainly matter at room temperature and are less pronounced at \ele.
\subsection{Application of the two-component coatings}
   \label{sec:2CC}

\begin{figure}[h]
                              \centering\begin{subfigure}[t]{.72\linewidth}
                              \centering
                              \includegraphics[width=\linewidth]{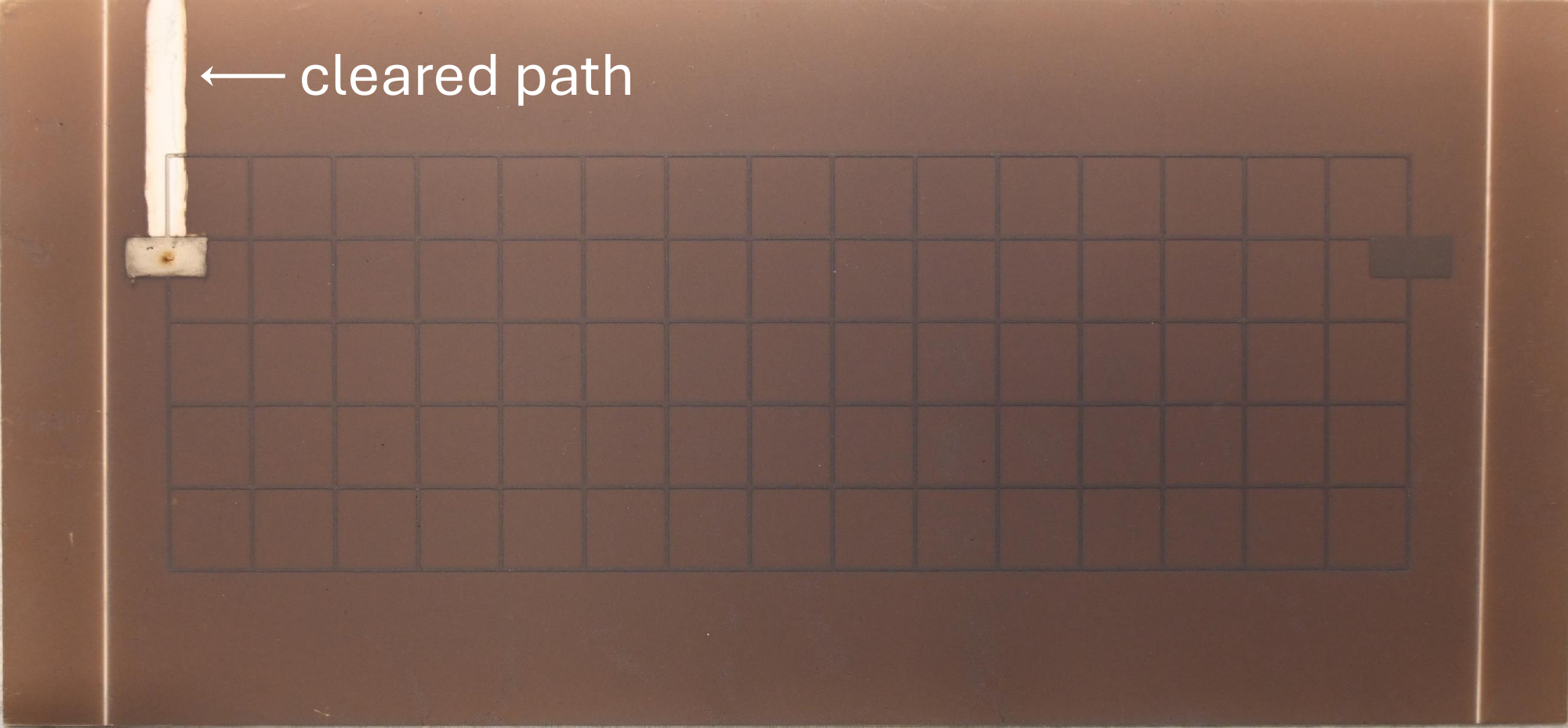}
                                      \pos{\tif{(a)}}
                                \label{fig:BTOMn_Mikroskopie_a}
                              \end{subfigure}
                              \hfill \\[0.15cm]
                              \begin{subfigure}[t]{0.33\linewidth}
                              \includegraphics[width=\linewidth]{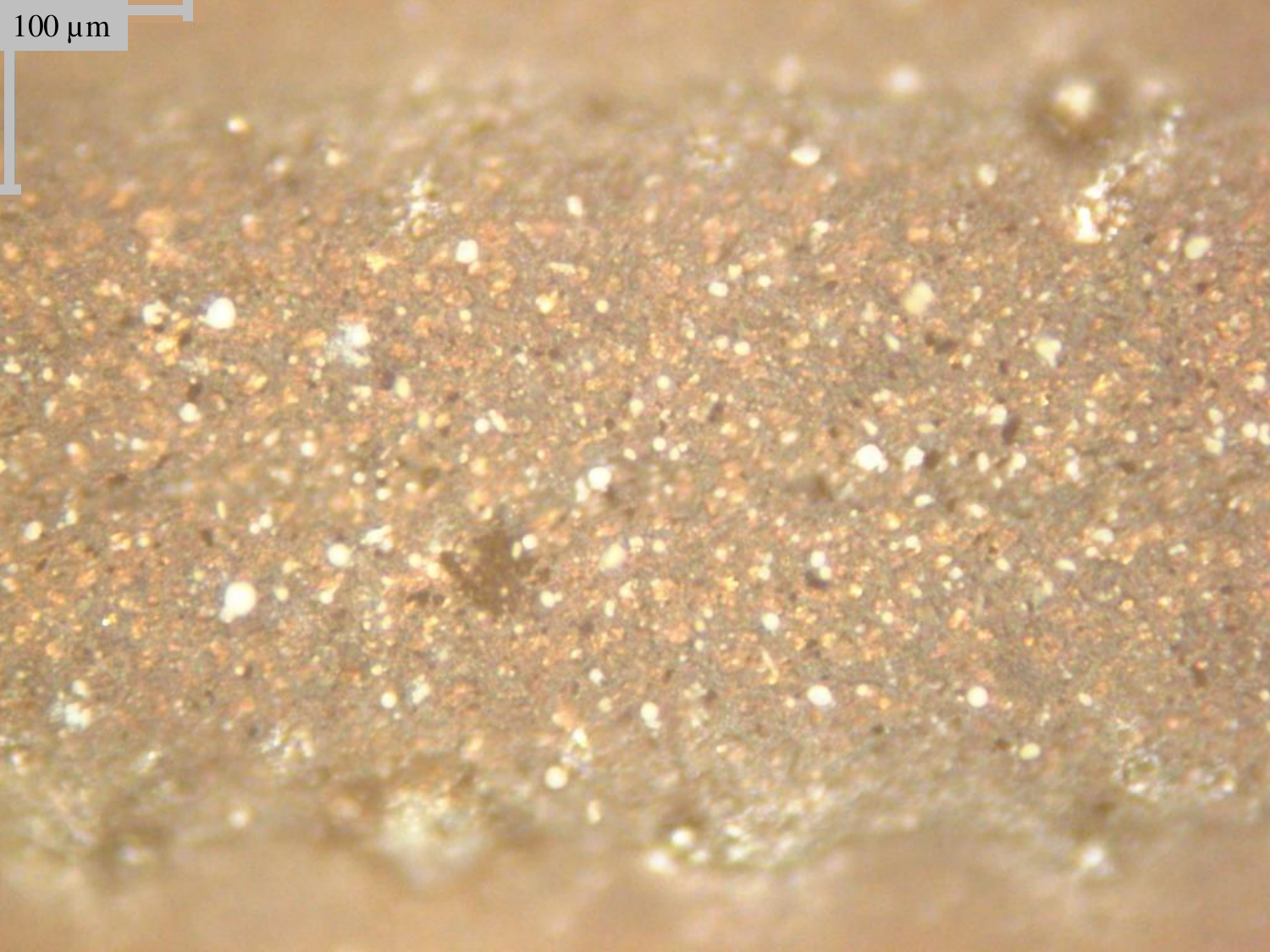}
                                      \pos{\tif{(b)}}
                                \label{fig:BTOMn_Mikroskopie_b}
                              \end{subfigure}
                              ~~~~
                              \begin{subfigure}[t]{0.33\linewidth}
                              \includegraphics[width=\textwidth]{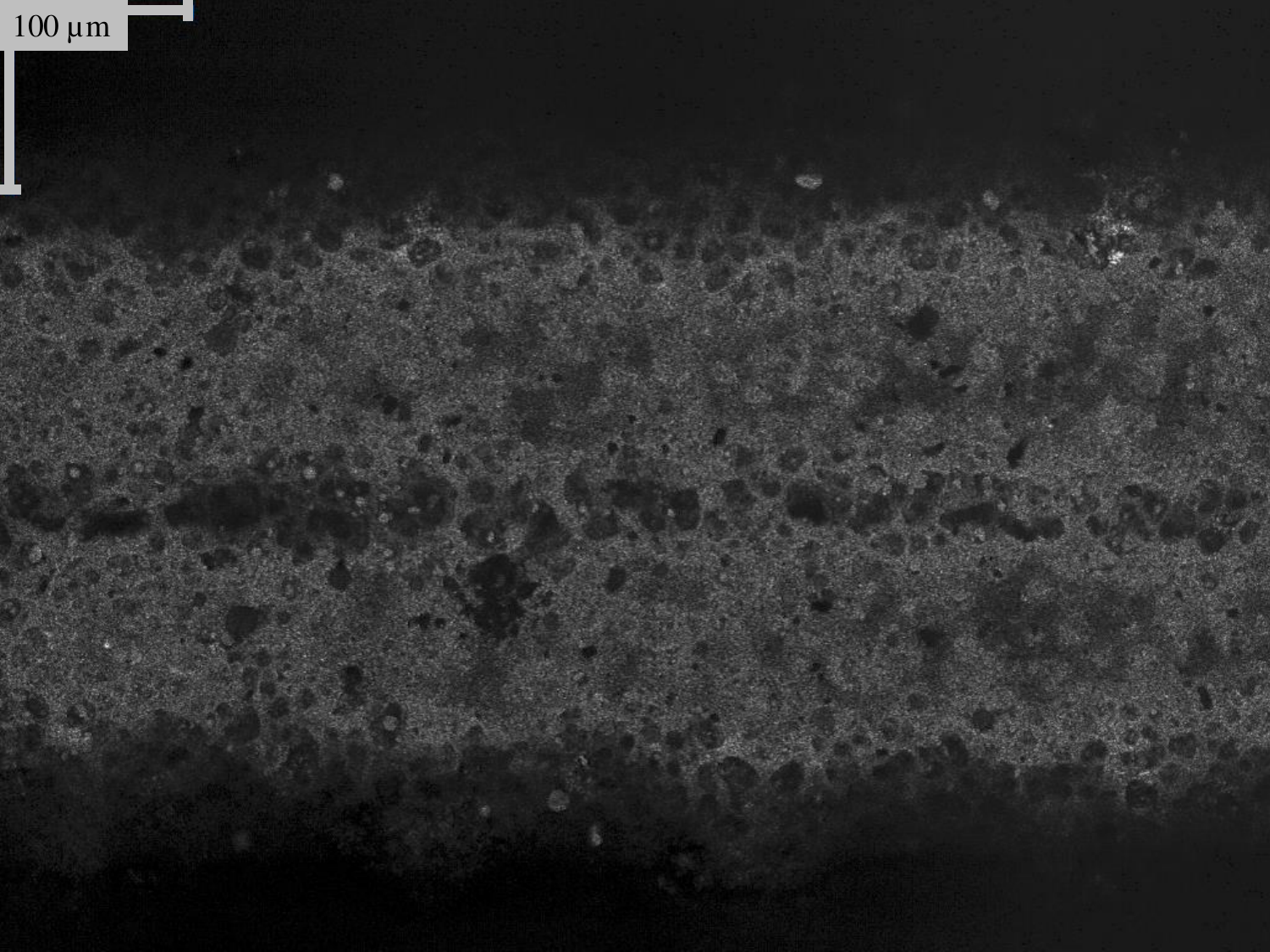}
                                      \pos{\tif{(c)}}
                                \label{fig:BTOMn_Mikroskopie_c}
                              \end{subfigure}
                              \caption{Photograph of an electrode configuration coated with a 1:1 ratio of BTO:catalyst and a theoretical loading of \SI{3}{mg.cm^{-2}} (a). Microscopy images show a single grid line at a 20x magnification in the optical mode (b) and laser mode (c).}
                              \label{fig:BTOMn_Mikroskopie}
                              \end{figure}

After the successful discharge ignition of fully BTO-coated electrode configurations and the subsequent assessment of $U_{\mathrm{pp}}$, $X$, $S$, and $CB$, the next step was to cover electrode configurations with two-component coatings and to repeat the assessment. 
Two electrode configurations, both with a total theoretical loading of \SI{3}{mg.cm^{-2}} and BTO:catalyst ratios of 1:1 and 1:2, respectively, were prepared.
An exemplary photograph depicting the 1:1 ratio coated electrode configuration is presented in Figure\,\ref{fig:BTOMn_Mikroskopie}a. 
The white uncoated area originates from the clearance of the path from the used contact points to the edge of the electrode configuration prior to any measurement to prevent coating abrasion and accumulation under the contact counterpart, which could potentially lead to irregular contact and the formation of an arc. 
The full-size image is directly followed by microscopy images of a single grid line of the same electrode to verify the full coating.
Noticeable, a coating structure resembling those observed in Figure\,\ref{fig:Mikroskopie}c can be seen upon close look, suggesting that BTO acts here in a structure-directing way.
In contrast, the color of the highly porous structure has turned dark, although some white irregularly embedded spheres of different sizes are still present in the voids of the structure.
Additionally, very few dark and light spots indicate local accumulation of one of the two components.
Nevertheless, the overall coating is rather homogeneous.
The laser microscopy image (see Figure\,\ref{fig:BTOMn_Mikroskopie}c) is intended to help discern the highly porous structure, yet it should be noted that the method is sensitive to differences in height. 
Thus, it is not possible to simultaneously focus on all levels and present the entire structure. 

\begin{figure}[h]
                              \centering
                              \includegraphics[trim={4.6cm 19.1cm 4.5cm 4.3cm},clip,width=\linewidth]{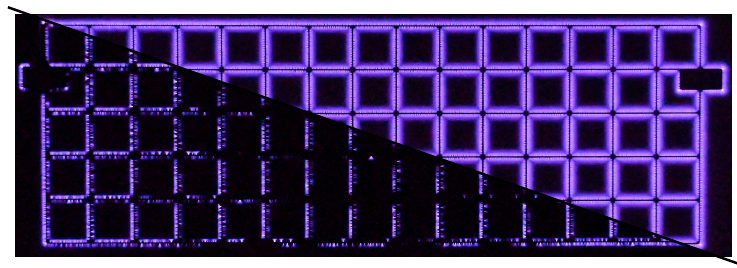}
                              \caption{Photographs of an ignited electrode configuration coated with a BTO:catalyst ratio of 1:1 and a theoretical loading of \SI{3}{mg.cm^{-2}}. The discharge was ignited in synthetic air at a peak-to-peak voltage of \SI{8.5}{kV} (below the diagonal) and \SI{10.5}{kV} (above the diagonal), respectively.}
                              \label{fig:BTOMn_ignition}
                              \end{figure}

Other than the pure BTO coated electrode configurations, the two-component coatings delay a fully homogeneous ignition due to the included catalyst and its discharge-inhibiting properties.
This becomes clear when comparing the illumination of the BTO:catalyst 1:1 coated electrode configuration at a peak-to-peak voltage of \SI{8.5}{kV} (see Figure~\ref{fig:BTOMn_ignition}) to those observed for the pure BTO coated ones shown in Figure\,\ref{fig:ignition} at the same voltage.
Although the regions near the contact points are nearly fully illuminated, several vacancies in the ignition pattern, appearing as darker spots, along the grid indicate a reduced number of streamers apparent as a more pronounced streamer separation.
Whether this separation is caused by a reduced temporal or spatial occurrence cannot be determined here due to the time-averaged nature of the photograph.
In addition, the illumination fades off toward the electrode configuration's center.
Applying higher peak-to-peak voltages resolves this issue as the electrode is more uniformly ignited.
To directly compare the illumination and discharge behavior, the two described voltages used for the ignition of the same electrode configuration with BTO:catalyst coating are shown in a combined way in Figure\,\ref{fig:BTOMn_ignition}.
Although already a slighter peak-to-peak voltage increase of less than \SI{0.5}{kV} would have been sufficient to ensure a more uniform illumination, the voltage of \SI{10.5}{kV} corresponds to the highest used in the ignition assessment and is therefore the brightest in the sequence, making it a better-suited choice for comparison.
In addition to the $\mo$ catalyst, pure \ch{MnO2} has been successfully tested as well regarding its ignition (not shown).

Since the full ignition of the electrode configuration covered with BTO and catalyst was found to be feasible, the plasma-assisted oxidation of $\nbu$ was investigated analogously to the BTO-coated electrode configurations. 

                              \begin{figure}[h]
                              \centering
                              \includegraphics[width=\linewidth]{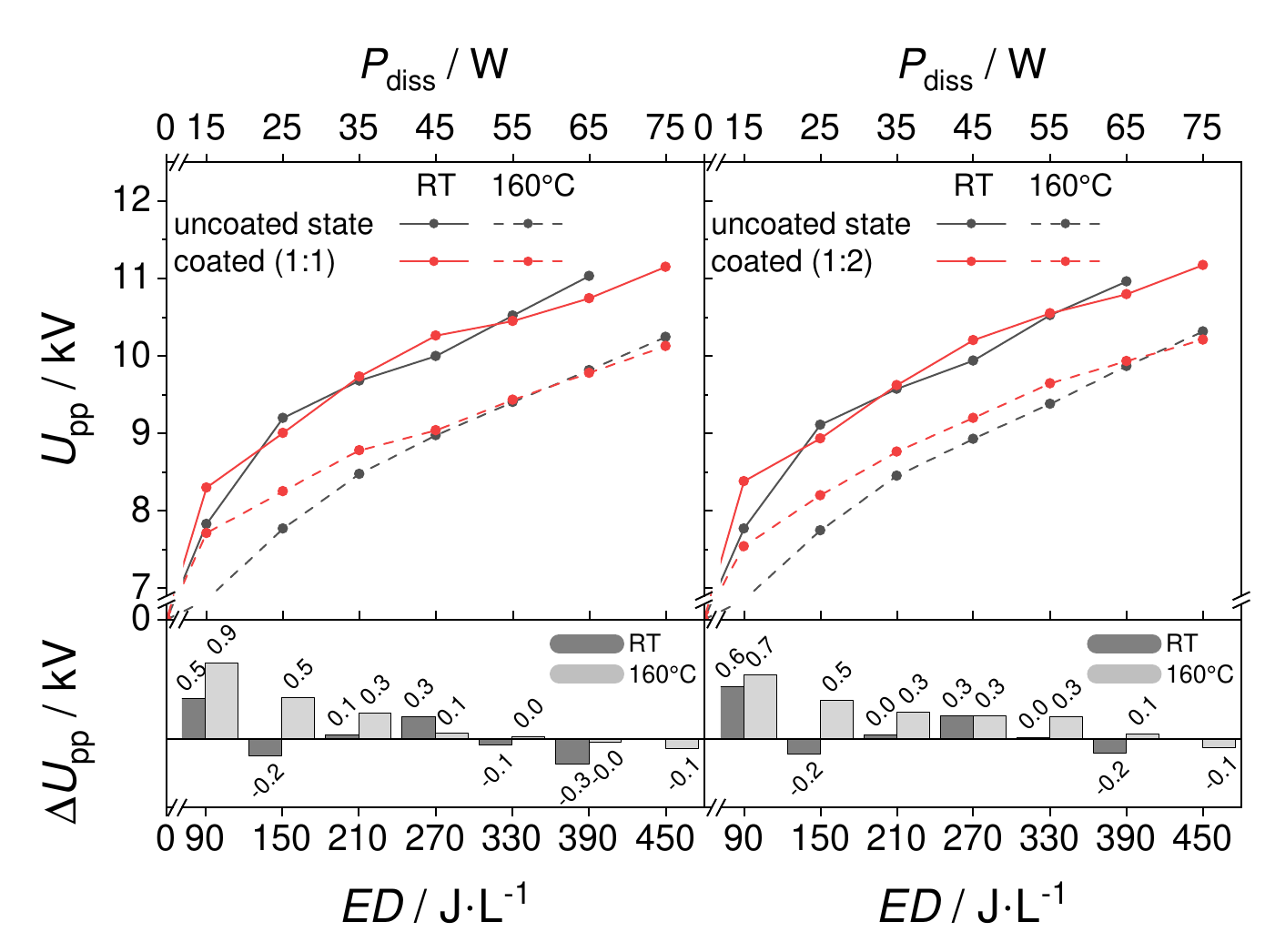}
                              \caption{Applied peak-to-peak voltages required to reach set values of dissipated power for both electrode configurations featuring two-component coatings with BTO:catalyst ratios of 1:1 and 1:2 and a theoretical loading of \SI{3}{mg.cm^{-2}}. An emphasis on the coating-induced changes is shown in the lower part, displaying $\Delta U_{\mathrm{pp}}$, the difference of the obtained values for the coated and uncoated state of each respective electrode configuration.}
                              \label{fig:BTOMn_Upp}
                              \end{figure} 

\begin{figure}[h]
                              \centering\includegraphics[width=\linewidth]{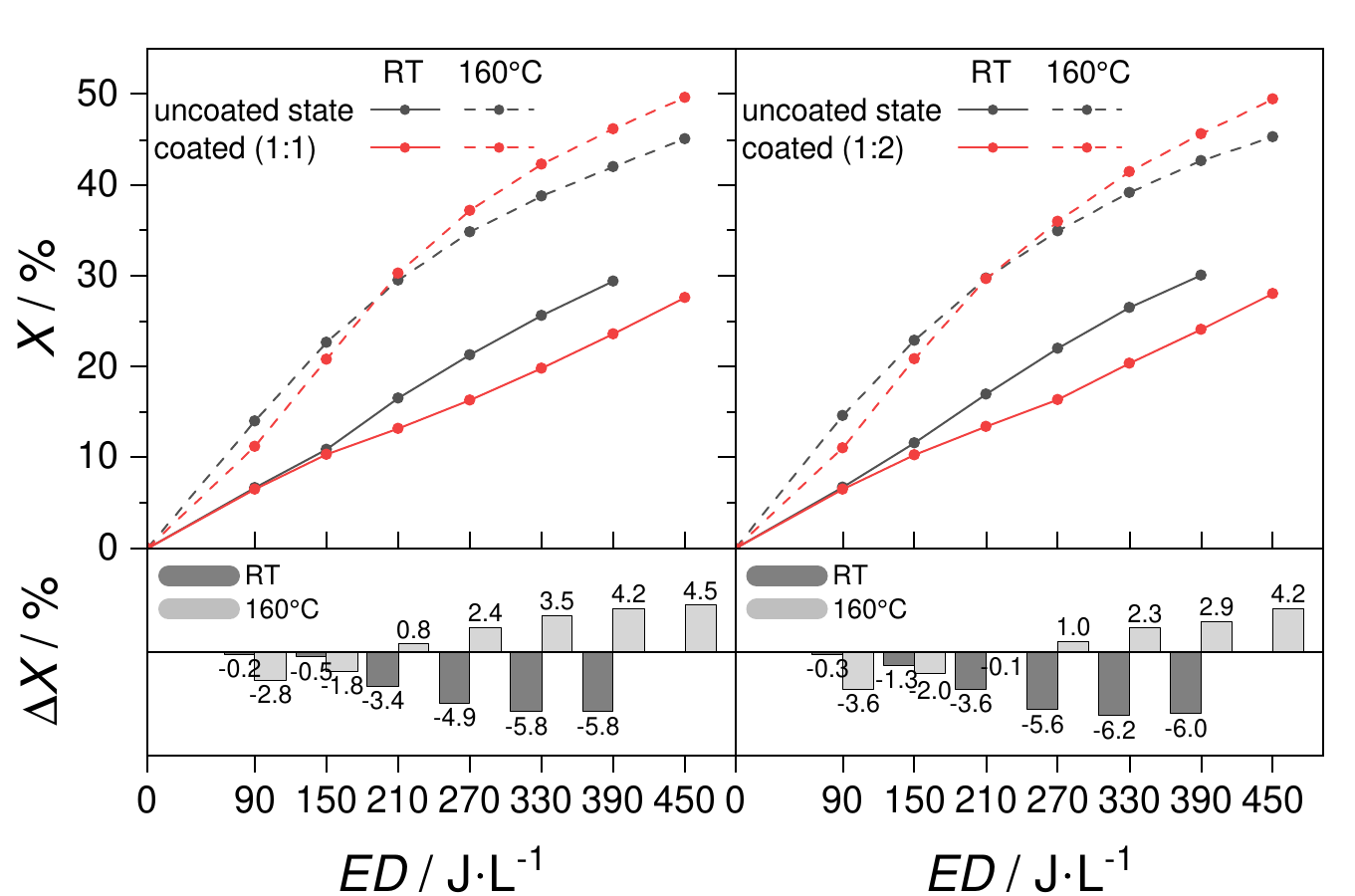}                              
                              \caption{Conversion achieved with the electrode configurations coated with BTO:catalyst ratios of 1:1 and 1:2 at room temperature (solid lines) and \ele ~ (dashed lines). The upper part of the diagram compares the coated (red) and the uncoated (black) state of the respective electrode configuration, while the lower part shows the difference.}
                              \label{fig:BTOMn_X}
                              \end{figure}

\begin{figure*}[h]
                              \centering\begin{subfigure}[t]{0.49\linewidth} 
                              \includegraphics[width=1.02\linewidth]{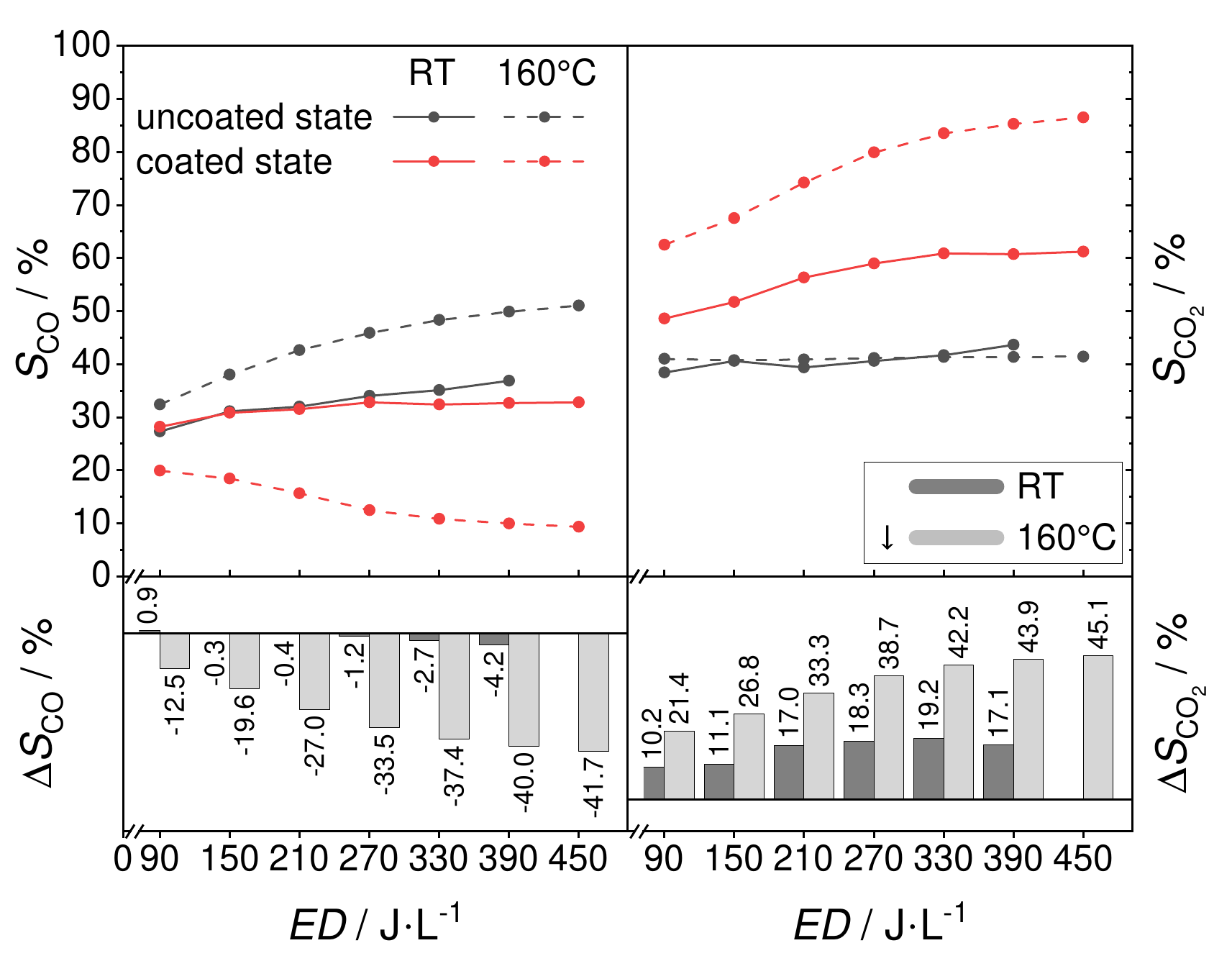}
                                      \put(-40,0){\tif{(a)}}
                                \label{BTOMn_S1}
                              \end{subfigure}
                              \hfill
                              \centering\begin{subfigure}[t]{0.49\linewidth} 
                              \includegraphics[width=1.02\linewidth]{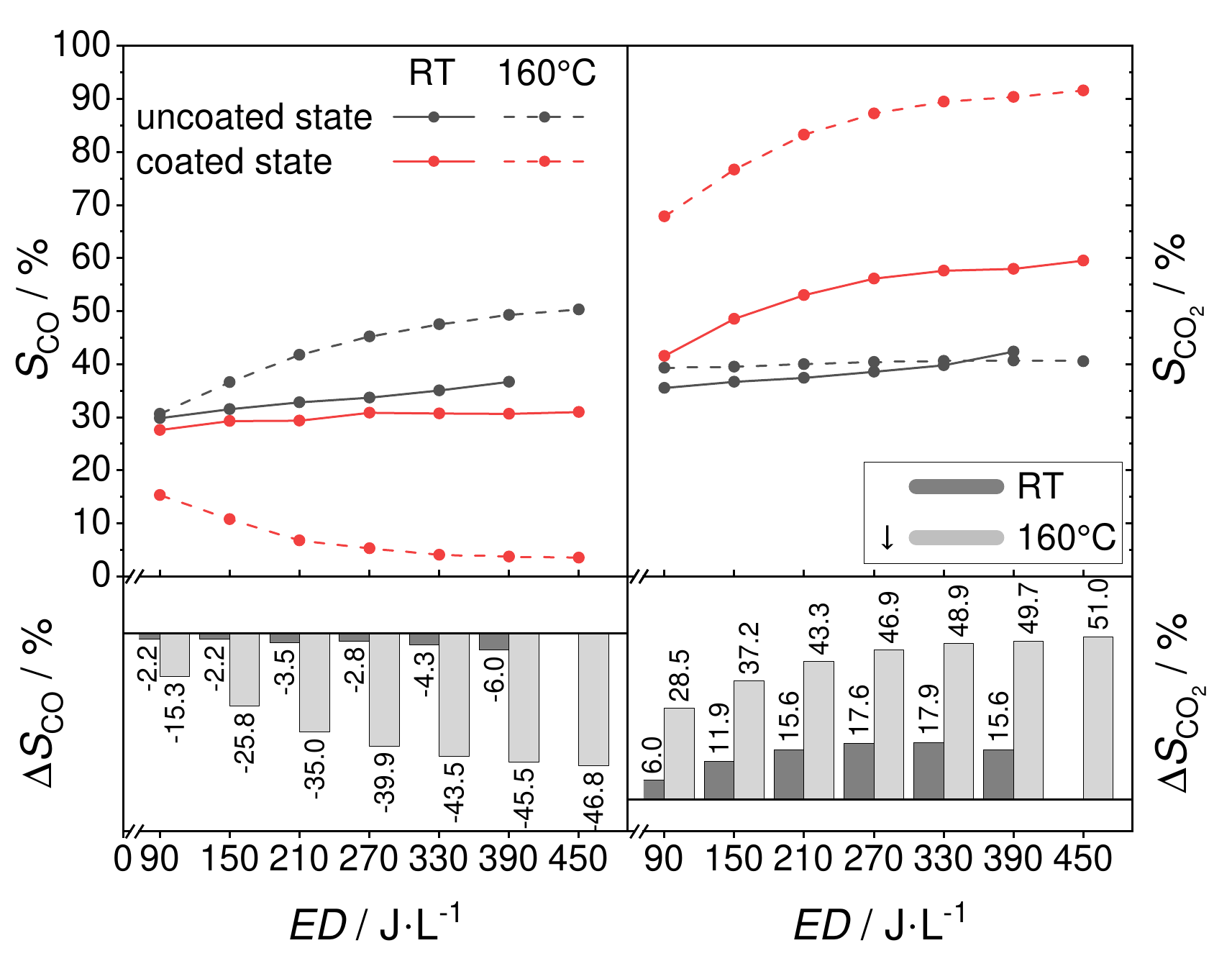}
                                      \put(-40,0){\tif{(b)}}
                                \label{BTOMn_S2}
                              \end{subfigure}                              
                              \caption{CO and \ch{CO2} selectivities for the electrode configurations coated with BTO:catalyst ratios of 1:1 (a) and 1:2 (b) at room temperature (solid lines) and \ele ~ (dashed lines). The upper part of the diagram compares the coated (red) and the uncoated (black) states of the respective electrode configuration, while the lower part shows the difference.}
                              \label{fig:BTOMn_S}
                              \end{figure*}

\begin{figure}[h]
                              \centering
                              \includegraphics[width=\linewidth]{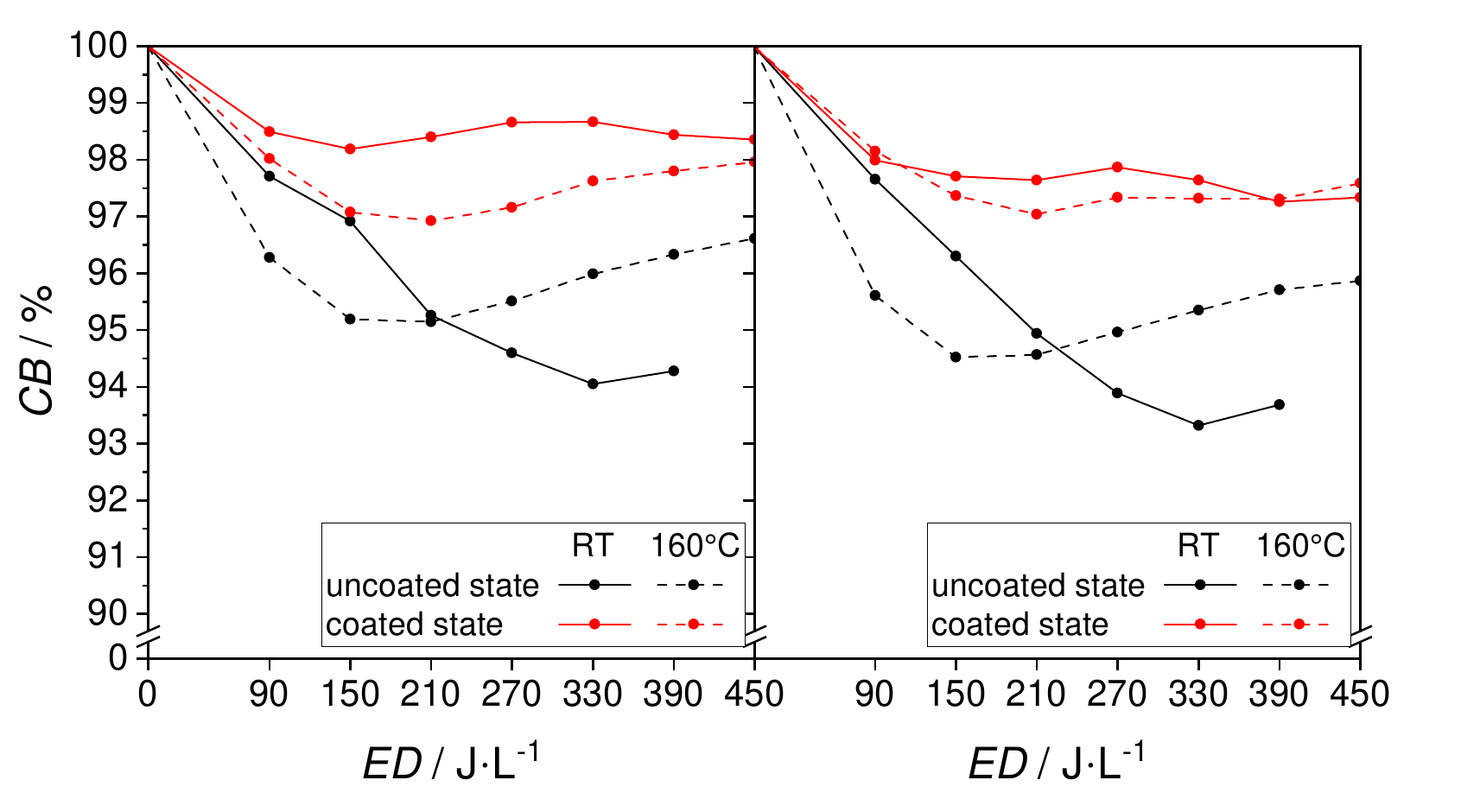}
                              \caption{Carbon balance for the two electrode configurations with BTO:catalyst ratios of 1:1 (left) and 1:2 (right) at room temperature (solid lines) and \ele ~ (dashed lines). The respective uncoated state is shown in black, whereas a red curve represents the coated state.}        
                              \label{fig:BTOMn_CB}
                              \end{figure}

Similar to Figure\,\ref{fig:Upp}, the additional peak-to-peak voltage application requirement due to the coating, $\Delta U_{\mathrm{pp}}$, diminishes with increasing power dissipation for the two-component coatings, as shown in Figure\,\ref{fig:BTOMn_Upp}.
One difference between the pure BTO coating and the two-component coatings is the more pronounced effect of temperature, which before led to only slight changes.
Here, the room temperature measurements result in $\Delta U_{\mathrm{pp}}$ fluctuating around zero, whereas heating to \ele ~ leads to a more coherently decreasing trend.
Notably, and despite the discharge ignition-inhibiting nature of the dielectric catalyst, the additional voltage requirement for both two-component coatings is comparable to the one determined for the higher purely BTO-coated electrode configuration.
Overall, the maximal applied peak-to-peak voltages for the two coatings are \SI{11.1}{kV} and \SI{11.2}{kV} at room temperature, and \SI{10.1}{kV} and \SI{10.2}{kV} at \ele ~ for the coating ratios of 1:1, and 1:2, respectively.
\\ The catalytic parameters were determined and are shown in Figures\,\ref{fig:BTOMn_X}--\ref{fig:BTOMn_CB}.
Comparison of different BTO:catalyst ratios exhibits only small changes with respect to $\nbu$ conversion.
In both cases, the presence of the coating led to a progressively lower conversion with rising energy density at room temperature, but to the opposite trend when heated to \ele. 
Since this temperature dependence has not been observed in such a pronounced way before, as presented in Figure\,\ref{fig:BTO_X}, explanations are likely to originate from the presence of the catalyst. 

Without the two-component coating, the highest achieved degrees of conversion are \SI{29.4}{\%} and \SI{45.1}{\%} at room temperature and at \ele, respectively, for the first electrode configuration, and \SI{30.1}{\%} and \SI{45.3}{\%} at room temperature and at \ele, respectively, for the second electrode configuration.
These values change to \SI{23.6}{\%} and \SI{49.6}{\%}, and to \SI{24.1}{\%} and \SI{49.5}{\%} at room temperature and at \ele, respectively, for the two coated states in the same order and at the same energy densities.

Besides conversion, selectivities were even more strongly influenced by the two-component coatings.
For both BTO:catalyst ratios, the amount of formed CO is only slightly reduced at room temperature with respect to the uncoated electrode configuration.
However, heating to \ele ~ leads to a significant decrease in CO formation that further proceeds with increasing energy density.
CO selectivity for a coated state decreases at \ele ~ from the lowest to the highest energy density from \SI{20.0}{\%} to \SI{9.4}{\%} for the 1:1 ratio and from \SI{15.4}{\%} to \SI{3.5}{\%} for the 1:2 ratio.
\ch{CO2}, in contrast, shows a substantial increase in formation already at room temperature over the energy density range from \SI{48.6}{\%} to \SI{61.2}{\%} and \SI{41.5}{\%} to \SI{59.5}{\%} for the two ratios of 1:1 and 1:2, respectively, corresponding to enhancements with respect to the uncoated state of \SI{10.2}{\%} to \SI{17.1}{\%}, and \SI{6.0}{\%} to \SI{15.2}{\%}, respectively.
These selectivity enhancements are further intensified at elevated temperature and then range from \SI{21.4}{\%} to \SI{45.1}{\%}, and from \SI{28.5}{\%} to \SI{51.0}{\%}, respectively, leading to a maximal \ch{CO2} selectivity of \SI{86.5}{\%} for the 1:1 ratio and of \SI{91.6}{\%} for the 1:2 ratio.
\\ Although in opposite directions, both CO and \ch{CO2} selectivities exhibit a flattening curve shape with respect to the uncoated state.
Therefore, lower CO and higher \ch{CO2} values can only be reached by increasing the catalyst content but not by further increasing the energy density.

As before, the $CB$ was determined according to Equation\,\ref{eq:CB} to assess the completeness of the species identification.
Figure\,\ref{fig:BTOMn_CB} shows a clear enhancement of the $CB$ found for the two-component coatings compared with the pure BTO coatings, as values above \SI{97}{\%} are reached for the total energy density range, and only the uncoated states exhibit lower values.
Thus, the addition of the catalyst into the BTO coating substantially lowered the amount of products originating from incomplete oxidation or different reaction pathways.

Generally, closer distances between the discharge region and coating are coupled to stronger interactions between these two areas.
This IPC configuration, which has been achieved in this work by two-component full coatings is, however, still a form of PPC configuration: the coating structures observed in this work resemble a highly porous network with small voids to allow for discharge ignition.
This means that the two more or less independent processes plasma treatment and catalysis still coexist, but in very close proximity, resulting in highly efficient catalytic CO oxidation to \ch{CO2} at \ele.
VOCs are rather activated and oxidized by and in the discharge and then further oxidized by the coating, drastically boosting the \ch{CO2} selectivity without significantly changing conversion.

\section{Discussion}
   \label{sec:discussion}

As shown in Figure\,\ref{fig:BTOMn_X}, the integration of the catalyst led to an increase in conversion at \ele.
To better compare the changes, Table\,\ref{tab:X} lists the degrees of conversion for a fixed energy density of \SI{390}{J.L^{-1}} at room temperature and \ele ~ for the BTO and the two BTO:catalyst coatings. 
It is evident that temperature has the most dominant effect, while the application of a pure BTO coating leads to very slight changes.
The integration of the catalyst into the coating leads to a decrease in conversion at room temperature, but to an increase at \ele, further amplifying the temperature influence. 

\begin{table*}[ht]
			\begin{center}
\caption{Summarized degrees of conversion and \ch{CO2} selectivities for the electrode configurations at \SI{390}{J.L^{-1}}. Only coatings with a theoretical loading of \SI{3}{mg.cm^{-2}} are included. Values listed for the uncoated state correspond to the mean value of the respective uncoated states of the electrodes later coated with BTO and BTO with catalyst.}
			\begin{tabular}[H]{cccccc}
			\toprule
         && \textbf{uncoated} & \textbf{BTO} & \multicolumn{2}{c}{\textbf{BTO:catalyst}} \\
          &          &                &                    &      1:1      &     1:2    \\
			\midrule
conversion ($X$)&
\makecell{RT\\ \ele}  &
\makecell{\SI{30.0}{\%} \\ \SI{41.8}{\%}}  &              
\makecell{\SI{31.7}{\%} \\ \SI{39.7}{\%}}  &              
\makecell{\SI{23.6}{\%} \\ \SI{46.2}{\%}}  &              
\makecell{\SI{24.1}{\%} \\ \SI{45.6}{\%}}  \\ \midrule 
\ch{CO2} selectivity ($S_{\COt}$)&
\makecell{RT\\ \ele}  & 
\makecell{\SI{42.3}{\%} \\ \SI{41.2}{\%}}  &               
\makecell{\SI{49.9}{\%} \\ \SI{40.0}{\%}}  &               
\makecell{\SI{60.7}{\%} \\ \SI{85.3}{\%}}  &               
\makecell{\SI{58.0}{\%} \\ \SI{90.4}{\%}}  \\               
                \bottomrule\label{tab:X}
			\end{tabular}
			\end{center}
			\end{table*}

Possible explanations involve either an unlikely interaction of the catalyst with $\nbu$, which is not able to catalyze $\nbu$ oxidation at \ele, or, more likely, with \ch{O2}.
An increasing supply of atomic oxygen or other highly reactive oxygen species by splitting of ozone is rather plausible.
Ozone is known to be formed in considerable amounts during plasma treatments in \ch{N2}/\ch{O2} gas mixtures\cite{Kogelschatz,O3_SDBD_VDBD}.
Simultaneously, \ch{MnO2} has been proven to be capable of splitting ozone catalytically \cite{O3Abbau1,O3Abbau2,futamura}, which is also not stable at elevated temperatures \cite{Wiberg}.
This relation and its consequences need to be addressed for verification in further work comprising generation and consumption measurements of both ozone and atomic oxygen during the plasma-assisted oxidation of alkanes. \\ 
The most intense influence was observed for the selectivities, as an increase of up to \SI{51.0}{\%} compared with the uncoated state was observed for the 1:2 ratio of the BTO:catalyst coating at \SI{450}{J.L^{-1}} and \ele.
This \ch{CO2} selectivity increase is generally coupled to a lower CO share, because formed CO is subsequently converted catalytically into \ch{CO2} by \ch{MnO2}.
An overall sequence of $\nbu$ splitting in the discharge region and consecutive oxidation of the partially oxidized intermediate species to \ch{CO2} as the final product is still considered most reasonable just as proposed in our previous work\cite{Niklas}.
The main difference is the optimized contact between plasma and catalyst due to the two-component full coating also covering the grid lines.
Although formally changing the configuration from PPC to IPC in this work, there is still no indication of any significant synergetic effects between discharge and catalyst despite the optimized proximity using the full two-component coating.
Although formally in the IPC configuration, the consecutive reaction and interaction schemes of PPC are more likely to apply in our case for the plasma-catalytic oxidation of $\nbu$ using a \ch{MnO2}-based catalyst.

\section{Conclusion}
   \label{sec:conclusions}
In this work, the plasma-driven oxidation of $\nbu$ as a model VOC was investigated using a twin SDBD as a plasma source.
Unlike electrode configurations fully coated with catalysts, which failed to ignite due to the discharge ignition-inhibiting properties of the applied coating, a uniform ignition along the electrode grid was achieved here by adding BTO, which was found to have a very low impact on the overall ignition and chemical processes.
Higher amounts of catalyst exhibit higher potential for better process efficiency than further increasing dissipated power.
The main benefit of using the catalyst was a significantly shifted selectivity in favor of \ch{CO2} while lowering the amount of formed CO and other by-products originating from incomplete oxidation or different pathways. 
In this way, a maximum selectivity to \ch{CO2} of \SI{91.6}{\%} was reached at \ele ~ for the 1:2 BTO:catalyst coating while simultaneously reducing CO selectivity down to \SI{3.5}{\%}. 
Compared with an uncoated electrode configuration, this equals a change of +\SI{51.0}{\%} for \ch{CO2} and -\SI{46.8}{\%} for CO.
Still, although using a formal IPC configuration with a maximized interaction between discharge and catalyst, the results do not point to a synergistic plasma--catalyst interaction for the used \ch{MnO2}-based catalyst.
Since the two-component coating approach was successfully established, the range for testing other materials, also those usually detrimental to the discharge ignition, with the SDBD electrode configuration was widened.
Continuation of this work with new materials may reveal a stronger interplay and potentially synergistic effects in the future.
   
\section{Acknowledgements}
   \label{sec:Acknowledgement}

The authors thank the Chair for Experimental Physics II of the Ruhr University Bochum for granting access to their laser scanning microscope. The investigation presented in this paper received financial support from the German Research Foundation (DFG) through projects A5 and A7 within the Collaborative Research Center SFB 1316 (project number 327886311), ``Transient atmospheric pressure plasmas - from plasmas to liquids to solids."

\section{Data availability statement}
The data that support the findings of this study are available at https://rdpcidat.rub.de/node/xxx.


\bibliography{lit.bib}%

\end{document}